\documentclass[12pt,a4paper]{amsart}

\usepackage{amsmath,amsfonts,amssymb,xy}

\newcommand{\C}{\mathbb C}

\def\d{\partial}
\def\CP1{\mathbb{C}\mathrm{P}^1}
\def\un{{1\!\! 1}}
\def\r{\mathfrak{r}}
\def\s{\mathfrak{s}}

\def\Res{\mathop{\mathrm{Res}}}

\def\shb{\sqrt\hbar}  
\def\End{\mathrm{End}}
\def\cV{\mathcal{V}}
\def\f{\mathfrak{f}}
\def\a{\mathfrak{a}}
\def\pr{(\r_\ell)}
\def\ps{(\s_\ell)}
\def\cF{\mathcal{F}}

\newtheorem{theorem}{Theorem}[section]

\newtheorem{lemma}[theorem]{Lemma}

\theoremstyle{definition}

\newtheorem{convention}[theorem]{Convention}

\title[Towards Lax for topological hierarchies]
{Towards Lax formulation of integrable hierarchies of topological type}

\author{G.~Carlet}

\address{G.~Carlet: 
Dipartimento di Matematica e Applicazioni \\
Universit\`a di Milano-Bicocca \\ 
Via R. Cozzi, 53 \\
20125 Milano \\
Italia}

\email{guido.carlet@unimib.it}

\author{J.~van de Leur}

\address{J.~van de Leur:
Mathematisch Insituut \\
Universiteit Utrecht \\
Postbus 80010 \\
3508 TA Utrecht \\
Nederland}

\email{j.w.vandeleur@uu.nl}

\author{H.~Posthuma}

\address{H.~Posthuma:
Korteweg-de Vries Instituut voor Wiskunde\\
Universiteit van Amsterdam \\
Postbus 94248 \\
1090 GE Amsterdam \\
Nederland}

\email{h.b.posthuma@uva.nl}

\author{S. Shadrin}

\address{S.~Shadrin:
Korteweg-de Vries Instituut voor Wiskunde\\
Universiteit van Amsterdam \\
Postbus 94248 \\
1090 GE Amsterdam \\
Nederland}

\email{s.shadrin@uva.nl}

\begin{document}

\begin{abstract}
To each partition function of cohomological field theory one can associate an Hamiltonian integrable hierarchy of topological type. The Givental group acts on such partition functions and consequently on the associated integrable hierarchies. We consider the Hirota and Lax formulations of the deformation of the hierarchy of $N$ copies of KdV obtained by an infinitesimal action of the Givental group. By first deforming the Hirota quadratic equations and then applying a fundamental lemma to express it in terms of pseudo-differential operators, we show that such deformed hierarchy admits an explicit Lax formulation.
We then compare the deformed Hamiltonians obtained from the Lax equations with the analogous formulas obtained in~\cite{BurPosSha1, BurPosSha2}, to find that they agree. We finally comment on the possibility of extending the Hirota and Lax formulation on the whole orbit of the Givental group action. 
\end{abstract}

\maketitle

\tableofcontents

\raggedbottom 

\section{Introduction}

\subsection{CohFTs and hierarchies}
A cohomological field theory (CohFT) is a system of factorizable forms on the moduli spaces of curves. It serves as an axiomatic setting that captures essential algebraic structures behind Gromov-Witten theory. Meanwhile, there are more examples of CohFTs, for example, coming from quantum singularity theory in the sense of Fan-Jarvis-Ruan~\cite{FJR}. In genus $0$ the notion of a CohFT is equivalent to a solution of the WDVV equation and, under some extra assumptions, turns out to be equivalent to the notion of (a formal germ of) a Frobenius manifold.

To an arbitrary CohFT one can associate a hierarchy of PDEs in (infinite-dimensional) Hamiltonian form. It was first constructed by Dubrovin and Zhang in the semi-simple homogeneous case \cite{DubZha2}, and their construction gives, indeed, a bi-Hamiltonian hierarchy based solely on genus $0$ data with some extra assumptions on dispersive behavoir of the tau function. In a weaker form, this construction is revisited in the recent papers of Buryak et al.~\cite{BurPosSha1,BurPosSha2}, where a Hamiltonian hierarchy was constructed using topological properties of the moduli spaces of curves in all genera.

In some special situations, related to singularity theory, there is an alternative construction of integrable hierarchies due to Givental and Milanov~\cite{GivMil}. They define a hierarchy in terms of Hirota quadratic equations (HQEs), where the vertex operators are defined in terms of the period integrals of the Lefschetz thimbles of the corresponding singularity. Their constructions are always presented ad hoc, and we are interested to understand whether we can use their ideas in a wider range of examples. A separate question is whether we can identify, in some explicit way, their HQEs with the Hamiltonian equations of Dubrovin and Zhang. 

\subsection{Givental group action}
The main tool used for the analysis of the integrable hierarchies in both approaches is the Givental theory of the group action on cohomological field theories. In the Hamiltonian formulation, we have a well-defined action on the Hamiltonians, the Poisson bracket, and the equations of the hierarchy. The infinitesimal deformations that correspond to the action of the Lie algebra of the Givental group are written down explicitly in~\cite{BurPosSha1,BurPosSha2}.

In the case of HQEs of Givental and Milanov \cite{GivMil} (and also \cite{Milanov1,Milanov2,MilTse}), the situation is a bit subtle. It is clear how to conjugate vertex operators with the elements of the Givental group, and this is exactly what they do in order to prove that a particular formal tau function satisfies the HQEs that they present. They specify an element of the Givental group that takes the tau function of several copies of KdV to a particular tau function that they consider. Then they conjugate the explicitly written vertex operators with this element of the Givental group, and show that absence of singularities near the potentially singular points is equivalent to the HQEs  of (several copies of) the KdV hierarchy.

The problem is that this procedure doesn't work in the opposite direction. If we take the simplest vertex operators needed to present the KdV hierarchy in form of the HQEs and try to conjugate them with an element of the Givental group, we get some series whose coefficients are divergent infinite sums. 

\subsection{Infinitesimal deformations of several copies of KdV} In this paper we perform a first step towards the understanding of the connection between the two approaches mentioned above: the action of the Givental group on HQEs of different hierarchies and the comparison of these hierarchies with Dubrovin-Zhang hierarchies known in Hamiltonian form. Namely, we look at the infinitesimal deformations of several copies of the KdV hierarchy. 

The action of the Lie algebra of the Givental group on the HQEs of KdV is well-defined, that is, it doesn't lead to divergent infinite sums. 
We also have explicit formulas for the action of the same Lie algebra on the Hamiltonian form of KdV. We translate these two actions from different sides to the Lax formulation of KdV and identify them modulo the ideal of KdV equations. 

Of course, it is in some sense a very expected result, since we compare two hierarchies of PDEs (two deformations of several copies of KdV) that have a common solution, and, moreover, are constructed in terms of this common solution. However, the comparative analysis of these two deformations turns out to be quite non-trivial, and there is an actual difference since only one of these tangent actions integrates to a group action.

We hope that our explicit computation can also give an idea of what kind of ``renormalization'' should be applied in order to have a well-defined action of the Givental group on the Hirota quadratic formulation of the hierarchies.

The fact that the Hirota quadratic formulation of the integrable hierarchies does not extend to arbitrary hierarchies associated to CohFTs is not very surprising from the point of view of the usual theory of integrable equations. On one hand the Dubrovin-Zhang construction \cite{DubZha2} (and the weaker construction from CohFTs of~\cite{BurPosSha1,BurPosSha2}) produces families of integrable hierarchies with good properties (bi-Hamiltonian structure, existence of tau function) which depend on a large number of parameters (e.g. the charge $d$ parametrizes the conformal $2$-dimensional Frobenius manifolds and therefore the corresponding Dubrovin-Zhang hierarchies with two dependent variables). On the other hand HQEs are only known for a much smaller subset of explicitly known hierarchies, typically those obtained as reductions of KP and 2D Toda type hierarchies. By ``explicitly known'' we mean hierarchies which can be represented in Lax form, which is usually given in terms of pseudo-differential or difference operators. 

A natural question therefore arises, if it is possible to extend not only the Hirota, but also the Lax formulation to more general families of integrable hierarchies, e.g. to CohFT hierarchies on the orbit of the Givental group action containing $N$-copies of KdV.

Our results, while only concerned with the infinitesimal deformations, demonstrate some progress in this direction and show several interesting features. First, we prove that it is indeed possible to express the deformed equations in Lax (or, equivalently Sato-Wilson) form using pseudo-differential operators. Second, while most examples of integrable hierarchies for which the Lax formulation is known are written in terms of a single Lax operator, our deformed equations feature $N$ Lax operators. Third, while we start by deforming second order differential operators of KdV type, the deformed operators are pseudo-differential operators where the non-trivial integral part is completely determined by the differential part. This suggests to view the deformed Lax equations as a sort of reduction of a deformation of multiple copies of the KP hierarchy. 

We expect that this approach will be very helpful in the construction of the Lax pairs and HQEs formulations for at least some subclass of the Dubrovin-Zhang hierarchies. 

\subsection{Organization of the paper} In section~\ref{KdV} we recall some basic facts about the KdV hierarchy and its Lax and Hirota formulations. In particular we review in detail how to recover, using a fundamental lemma, the Sato-Wilson and Lax equations from the Hirota bilinear equations, first in the case of the single KdV hierarchy and than in the case of $n$-copies of KdV. In section 2 we compute the deformations of the vertex operators induced by the infinitesimal Givental action of the twisted loop group. We also comment on the possibility of computing the global Givental group action on the vertex operators. Explicit formulas for the deformed Sato-Wilson and Lax equations are obtained in section 4 by a careful application of the fundamental lemma to the deformed Hirota equations. Finally, in section 5, the deformation formulas for Hamiltonians, recovered as residues of the Lax operators, are shown to coincide with those obtained in~\cite{BurPosSha1}.

\subsection{Acknowledgement}
G.C. wishes to acknowledge the support of the GQT cluster and of  ESF - MISGAM grants. 
S.S. was supported by a Vidi grant of the Netherlands Organization for Scientific Research.


\section{The KdV hierarchy}
\label{KdV}
We briefly recall some basic constructions in the theory of the KdV hierarchy using the pseudo-differential operator formalism. First we sketch the definition of Lax, Sato-Wilson and Hirota quadratic equations, omitting most proofs. Then we give a more careful treatment of the derivation of Lax and Sato-Wilson equations from the Hirota equations, since this will be required later. 

For more details see e.g.~\cite{Dic}.

%
%
%
%
%
%
%
%

\subsection{From Lax to Hirota}

The KdV hierarchy is a sequence of commuting flows in Lax form
\begin{equation}
\label{Lax}
\frac{\partial L}{\partial q_n}=\frac{a_n}{\sqrt\hbar}\left[\left(L^{n+\frac12}\right)_+,L\right]\, , \qquad n =0,1,2,\dots\, 
\end{equation} 
on the space of Lax operators 
\begin{align}\label{eq:Lax-op}
L= \hbar\partial^2+2u(x) \qquad \mbox{with }\partial:=\frac{\partial}{\partial x}\, .
\end{align}
Here $u(x)$ will be seen as a formal power series in $\shb$, i.e. $u(x;\shb) = u_0(x) + u_1(x) \shb + \dots$


Note that a solution $L=L(x,q)$ will be a function of $x$ and of all the times of the hierarchy $q=(q_0, q_1,\dots)$.

The pseudo-differential operator $L^{\frac12}$ is the square root of $L$ (i.e. the unique operator $L^\frac12 = \shb\partial+\dots$ such that $(L^\frac12)^2 = L$) and is equal to
\begin{align}
L^{\frac12}=& \sqrt{\hbar}\partial+\frac{u}{\sqrt{\hbar}}\partial^{-1}
-\frac{u'}{2 \sqrt\hbar
  }\partial^{-2} \\ \notag
& +\frac{\hbar u''-2 u^2}{4 \hbar^{3/2}}\partial^{-3}
-\frac{\hbar u^{(3)}-12 u u'}{8 \hbar^{3/2}}\partial^{-4}\\ \notag
& +\frac{\hbar^2 u^{(4)}-28 \hbar u u''-22 \hbar \left(u'\right)^2+8 u^3}{16
   \hbar^{5/2}}\partial^{-5}+\cdots\, .
\end{align}

Given a pseudo-differential operator $A$, i.e. a Laurent series $A = \sum_{n=-\infty}^m a_n(x) \partial^n$ for arbitrary $m$, we denote its differential part by $A_+:=\sum_{n=0}^m a_n(x) \partial^n$, while $A_- := A-A_+$. The product of pseudo-differential operators is defined by
\begin{equation}
\partial^k f(x) = \sum_{l=0}^\infty \binom{k}{l} \frac{\partial^k f}{\partial x^k} \partial^{k-l}.
\end{equation}

By $a_n$, $n=0,1,2,\dots$, we denote the constants
\begin{equation}
a_n=\frac{1}{(2n+1)!!}\, .
\end{equation}

The commutativity of the flows defined by the Lax equations~\eqref{Lax} follows from the so-called Za\-kha\-rov-Shabat (or zero curvature) equations
\begin{equation}
\label{ZS}
\frac{\sqrt\hbar}{a_k}\frac{\partial \left(L^{\ell+\frac12}\right)_+}{\partial q_k} 
-\frac{\sqrt\hbar}{a_\ell}\frac{\partial\left( L^{k+\frac12}\right)_+}{\partial q_\ell}=
\left[ \left( L^{k+\frac12}\right)_+,\left( L^{\ell+\frac12}\right)_+\right]\, .
\end{equation} 
Note that in this equation we can change all projectors $(\cdot)_+$ to $-(\cdot)_-$.


To a solution $L$ to the Lax equations~\eqref{Lax} one can associate a dressing operator, i.e. a pseudo-differential operator of the form
\begin{equation} \label{Pkdv}
P(\shb\partial)=P(x,q, \shb\partial) = 1 + p_1(x,q) (\shb\partial)^{-1} + \dots 
\end{equation}
such that
\begin{equation} \label{LPP}
L = P(\shb\partial)\hbar \partial^2 P(\shb\partial)^{-1}
\end{equation}
which satisfies the Sato-Wilson equations for the KdV hierarchy 
\begin{equation} 
\frac{\partial P(\shb\partial)}{\partial q_n} = -\frac{a_n}{\sqrt\hbar} (P(\shb\partial) (\shb\partial)^{2n+1} P(\shb\partial)^{-1} )_- P(\shb\partial)   \label{sato-kdv} .
\end{equation}

On the other hand a solution~\eqref{Pkdv} to Sato-Wilson equations~\eqref{sato-kdv} which satisfies the constraint
\begin{equation}
(P(\shb\partial) \hbar\partial^2 P(\shb\partial)^{-1})_- = 0 
\end{equation}
defines, through~\eqref{LPP}, a solution $L$ to the KdV hierarchy. 


Let us define the vertex operator $\Gamma$ as
\begin{equation} \label{S0}
\Gamma(q,\lambda) = \Gamma_+(q,\lambda) \Gamma_-(q,\lambda)
\end{equation}
where
\begin{align}
\label{S} 
\Gamma_- ( q, \lambda)&=
\exp\left( - {\sqrt\hbar}\sum_{n=0}^\infty b_n\lambda^{-2n-1}\frac{\partial}{\partial q_{n}}\right) \, ,
\\ 
\Gamma_+ (q, \lambda)&=\exp \left(  \frac{1}{\sqrt\hbar}\sum_{n=0}^\infty a_n \lambda^{2n+1}q_{n}\right) \, . \label{S1}
\end{align}
Here, the constants $b_n$ are given by
\begin{equation}
b_n:=(2n-1)!! \quad n=0,1,\dots 
\end{equation}

The wave function is defined as
\begin{equation}
\label{wave2}
\psi(x,q ,\lambda)=P (x,q, \lambda)\Gamma_+ ( q,\lambda)e^{\frac{x\lambda}{\sqrt\hbar}}\, .
\end{equation}
Note that here we have denoted by $P(x,q,\lambda)$ the (right) symbol of the dressing operator~\eqref{Pkdv}
\begin{equation}
P(\lambda)=P(x,q,\lambda) = 1+ p_1(x,q) \lambda^{-1} + \dots .
\end{equation}
The wave function satisfies the following linear system
\begin{align} \label{lin}
L\psi(x,q,\lambda)&=\lambda^2\psi(x,q,\lambda)\, ,\\ \notag
\frac{\partial \psi(x,q,\lambda)}{\partial q_n}&=\frac{a_n}{\sqrt\hbar}\left( L^{n+\frac12}\right)_+\left(\psi(x,q,\lambda)\right)\, .
\end{align}

It is well known that the wave function also satisfies the following system of quadratic equations \begin{equation}
\label{bil}
\Res_{\lambda}\lambda^{2p}\psi(x,q,\lambda) \psi(x',q',-\lambda) \, d\lambda=0,\quad p=0,1,2,\ldots,
\end{equation}
where $\Res_{\lambda}\sum_i a_i\lambda^i\, d\lambda=a_{-1}$. 


The KdV hierarchy can also be expressed in terms of the tau function. In particular one can prove that  for any solution $P$ to the the Sato-Wilson equations~\eqref{sato-kdv} there exists a function $\tau=\tau(x,q)$ such that
\begin{equation}
P(x,q,\lambda) = \frac{\Gamma_-(q, \lambda) \tau(x,q)}{\tau(x,q)} .
\end{equation}
Note that the tau function is uniquely determined by a solution of the Sato-Wilson equation up to a multiplicative constant. 


The wave function~\eqref{wave2} is given in terms of the tau function by
\begin{equation} \label{wave}
\psi(x,q,\lambda)=\frac{\Gamma (q,\lambda)\tau(x,q)}{\tau(x,q)}e^{\frac{x\lambda}{\sqrt\hbar}} \, .
\end{equation}

%

It is obvious from the quadratic equations for the wave function~\eqref{bil} that the tau function also satisfies similar quadratic equations, viz.,
\begin{equation} 
\Res_{\lambda}\lambda^{2p} (\Gamma(q,\lambda)\tau(x,q) ) \,  (\Gamma(q',-\lambda)\tau(x',q'))\, d\lambda=0,\quad p=0,1,2,\dots .
\end{equation}
These are called Hirota quadratic equations for the tau function. We proceed to consider them in more detail. 

\subsection{Hirota quadratic equations} 
The Hirota quadratic equations (HQEs) for the KdV hierarchy can be written in compact form as
\begin{equation}~\label{bil-com}
\Res_\lambda \lambda^{2p}\, \big(\Gamma(\lambda) \otimes \Gamma(-\lambda) \big) \, ( \tau \otimes \tau ) \,d\lambda =0
\quad p \geq 0.
\end{equation}�
The HQEs encode an infinite number of quadratic relations for the function $\tau$ and its derivatives w.r.t. the variables $q=(q_0, q_1, \dots)$.

We will assume that the function $\tau(q)$ is a formal power series in the variables $q$ and, in particular, that it is of the form 
\begin{equation} \label{asympt}
\tau(q) = e^{ \hbar^{-1} \cF(q;\shb)}
\end{equation}
for  $\cF(q;\shb)$ a formal power series in $q$ and $\shb$.

Equation~\eqref{bil-com} is interpreted as follows. After evaluating the two factors of the tensor product in $q$ and $q'$, respectively, it is written as
\begin{equation} \label{biltau}
\Res_{\lambda}\lambda^{2p}\, \left(\Gamma(q,\lambda)\tau(q)\right) \, \left(\Gamma(q',-\lambda)\tau(q')\right)\, d\lambda=0,\quad p=0,1,2,\dots 
\end{equation}
The vertex operator $\Gamma(q,\lambda)$ has been defined in~\eqref{S0}-\eqref{S1}. 
Passing to the variables $\eta$, $\xi$ defined by
\begin{equation}
q_n = \xi_n + \eta_n, \quad
q'_n = \xi_n - \eta_n
\end{equation}
one rewrites the argument of the residue as $\lambda^{2p}$ multiplied by
\begin{equation} \label{resarg}
e^{\frac2{\sqrt\hbar} \sum_{n\geq0} a_n \lambda^{2n+1} \eta_n} e^{- \sqrt\hbar \sum_{n\geq0} b_n \lambda^{-2n-1} \frac{\partial}{\partial{\eta_n}}} \tau(\xi+\eta) \tau(\xi-\eta) .
\end{equation}
This is a formal power series in the variables $\eta$ with coefficients which are Laurent series in $\lambda^{-1}$, i.e., with upper bounded powers of $\lambda$. The HQEs~\eqref{bil-com} state that in each of these Laurent series the terms with odd negative powers of $\lambda$ are vanishing.

We can also express the Hirota equations in terms of regularity of the differential $1$-form
\begin{equation} \label{bil-tau-ave}
\left( \Gamma(\lambda)\otimes \Gamma(-\lambda) - \Gamma(-\lambda)\otimes \Gamma(\lambda) \right) (\tau\otimes\tau) \, d\lambda .
\end{equation}
We say that this $1$-form is regular in $\lambda$ if, after the change of variables and the expansion in $\eta$ as before, the resulting Laurent series have no polar part, i.e. they are polynomials in $\lambda$. Since the averaging in~\eqref{bil-tau-ave} kills exactly the even terms, this is equivalent to the vanishing of the odd negative powers of $\lambda$ in~\eqref{resarg}. Hence $\tau(q)$ satisfies the Hirota quadratic equations for KdV iff the $1$-form~\eqref{bil-tau-ave} is regular.

\subsection{A fundamental lemma}

To derive the Sato-Wilson equations from the HQEs we need the following Fundamental Lemma which will also be important in Section~\ref{Givental-Lee}. 

\begin{lemma}\emph{(Fundamental Lemma)}\label{fundament}
Let $P(x,\sqrt\hbar\partial)$ and $Q(x,\sqrt\hbar\partial)$ be two pseudo-differential operators. Then the equation
\begin{equation} \label{fl1}
\Res_\lambda
P(x,\lambda)e^{\frac{x\lambda}{\sqrt\hbar}}Q(x',-\lambda)e^{-\frac{x'\lambda}{\sqrt\hbar}}\, d\lambda=0
\end{equation}
is  equivalent to
\begin{equation} \label{fl2}
\left(P(x,\sqrt\hbar\partial)\cdot Q(x,\sqrt\hbar\partial)^*\right)_-=0\, .
\end{equation}
\end{lemma}

Here $P(x,\lambda)$ denotes the right symbol of $P(x, \shb\partial)$ and $Q^*$ is the formal adjoint of $Q$, defined by 
$(c(x)\partial^k)^* = (-\partial)^k\cdot c(x)$.

We give a short proof of this Lemma, along long the lines of~\cite{Dic} and~\cite{KacLeu}.

\begin{proof}
The proof is based on the following identity, which can be checked by direct computation
\begin{equation} \label{pr1}
\Res_\lambda P(x,\lambda) Q(x,-\lambda) \, d\lambda = 
\sqrt\hbar \Res_\partial P(x, \sqrt\hbar\partial ) \cdot Q(x,\sqrt\hbar\partial)^* .
\end{equation}

Equation~\eqref{fl2} is equivalent to 
\begin{equation}
\Res_\partial P(x, \shb\partial ) \cdot Q(x, \shb\partial)^* \cdot \partial^k = 0 \quad
\text{for each } k\geq0
\end{equation}
or, introducing an extra formal variable $y$, to the generating identity
\begin{equation} \label{pr2}
\Res_\partial P(x, \shb\partial) \cdot Q(x,\shb\partial)^* \cdot e^{- y \partial} = 0 .
\end{equation}
Noting that
\begin{align}
Q(x,\shb\partial)^* \cdot e^{- y \partial} 
&= \left( e^{y \partial} \cdot Q(x,\shb\partial) \right)^* \\
&= \left( Q(x+y, \shb\partial) \cdot e^{y \partial} \right)^*
\end{align}
and making use of equation~\eqref{pr1}, we can rewrite equation~\eqref{pr2} as
\begin{equation}
\Res_\lambda P(x,\lambda) Q(x+y, -\lambda) e^{-\frac{y \lambda}{\shb}}\, d\lambda=0 .
\end{equation}
After a change of variable $y=x'-x$, this is exactly equation~\eqref{fl1}.
\end{proof}

%
%
%
%
%
%
%

\subsection{From Hirota to Lax}

Let $\tau(q)$ be a solution to the Hirota quadratic equations~\eqref{biltau}. Let us introduce a dependence on $x$ (and $x'$) by shifting $q_0 \to q_0 + x$ and $q_0' \to q_0' +x'$. Denote $\tau(x,q) = \tau(q_0+x, q_1, \dots)$ and 
\begin{equation} \label{defP} 
P(x,q,\lambda) = \frac{\Gamma_-(q,\lambda) \tau(x,q)}{\tau(x,q)} .
\end{equation}
Substituting in~\eqref{biltau} and dividing by $\tau(x,q) \tau(x',q')$ we obtain
\begin{equation}
\Res_\lambda \Big( \lambda^{2p} \Gamma_+(q,\lambda) P(x,q,\lambda) e^{\frac{x}{\shb}\lambda}  \Gamma_+(q',-\lambda) P(x',q',-\lambda) e^{-\frac{x'}{\shb}\lambda} \Big)\, d\lambda = 0 .
\end{equation}

Using the Fundamental Lemma we rewrite the HQEs as the following bilinear equation involving pseudo-differential operators
\begin{equation}  \label{bil-dif}
\left(
P(x,q,\sqrt\hbar\partial) \Gamma_+(q,\sqrt\hbar\partial)
\cdot \hbar^p \partial^{2p} \cdot
\Gamma_+(q',\sqrt\hbar\partial)^* P(x,q',\sqrt\hbar\partial)^*
\right)_-=0\, .
\end{equation}

Let us examine some consequences of this equation. 

First, set $p=0$ and $q=q'$. Since $\Gamma_+(q,\sqrt\hbar\partial)^*=\Gamma_+(q,\sqrt\hbar\partial)^{-1}$, one finds
\begin{equation}
\left(P(x,q,\sqrt\hbar\partial) P(x,q,\sqrt\hbar\partial)^*\right)_-=0\, .
\end{equation}
This implies, together with the fact that, by the definition~\eqref{defP}, $P(x,q,\sqrt\hbar\partial)=1+\left(P(x,q,\sqrt\hbar\partial)\right)_-$, that 
\begin{equation}
P(x,q,\sqrt\hbar\partial)^*=P(x,q,\sqrt\hbar\partial)^{-1}\, .
\end{equation}

Second, for $p=1$ and $q=q'$ we have
\begin{equation}
\left(P(x,q,\sqrt\hbar\partial) \hbar\partial^2 P(x,q,\sqrt\hbar\partial)^{-1}\right)_-=0\, .
\end{equation}
This implies that the pseudo-differential operator defined by
\begin{equation}
\label{LaxP}
L=L(x,q,\shb\partial):=P(x,q,\sqrt\hbar\partial)\hbar\partial^2 P(x,q,\sqrt\hbar\partial)^{-1}\, ,
\end{equation}
is actually second order differential operator of the form
\begin{equation}
L=\hbar\partial^2+2u(x,q) .
\end{equation} 

This gives immediately that
\begin{equation}
\label{u}
u=\hbar\frac{\partial^2 \log\tau}{\partial x\partial q_0}\, .
\end{equation}

Third,  by differentiating equation~\eqref{bil-dif} w.r.t. $q_n$ and setting $q=q'$ (and $p=0$), since
\begin{equation}
\frac{\partial\Gamma_+(q, \shb\partial)}{\partial q_n} = \frac{a_n}{\shb} (\shb\partial)^{2n+1} \Gamma_+(q,\shb\partial),
\end{equation}
we obtain the Sato-Wilson equations
\begin{equation} \label{Sato}
 \frac{\partial P (\sqrt\hbar\partial)}{\partial q_n}P (\sqrt\hbar\partial)^{-1}
=-\frac{a_n}{\sqrt\hbar}\left(
P (\sqrt\hbar\partial) (\sqrt\hbar\partial)^{2n+1}P (\sqrt\hbar\partial)^{-1}\right)_-
\, .
\end{equation}
Here and in the following we omit explicit dependence on $x$, $q$ when clear from the context. 

As explained before, the Lax equations~\eqref{Lax} are an immediate consequence of the Sato-Wilson equations. 
Summarizing, we have proved that a tau function $\tau(q)$ which satisfies the HQEs  defines a solution $u(x,q)$ of the KdV hierarchy.

\subsection{The residues}

Here we collect some observations on the residues that will be needed later. Taking the residue of~\eqref{defP} we find
\begin{equation}
\label{resP}
\Res_\partial P(\sqrt\hbar\partial)=-\frac{\partial\log \tau}{\partial q_0}\, .
\end{equation}
If we use this together with the residue of the Sato-Wilson equation~\eqref{Sato}, we obtain
\begin{equation}
\label{resL}
\frac{a_n}{\sqrt\hbar} \Res_\partial L^{n+\frac12}= \frac{\partial^2\log \tau}{\partial q_0\partial q_n}\, .
\end{equation}
Differentiating this equation w.r.t. $q_m$ we obtain
%
\begin{equation}
\label{resL2}
\frac{a_ma_n}{\hbar} \Res_\partial \left[\left(L^{m+\frac12}\right)_+,\left(L^{n+\frac12}\right)_-\right]= \frac{\partial^3\log \tau}{\partial q_0\partial q_m\partial q_n}\, .
\end{equation}
Note that, since the dependence of $\tau$ on $x$, $q_0$ is only through $q_0 +x$, we can always replace the derivative w.r.t. $q_0$ with the derivative w.r.t. $x$ and vice-versa.


\subsection{Multiple copies of KdV}
By definition, the tau function of $N$ copies of the KdV hierarchy is the product 
\begin{equation} \label{prodtau}
\tau(q)=\prod_{\gamma=1}^N \tau_\gamma(q_{\gamma})\, ,
\end{equation}
where each $\tau_\gamma(q_\gamma)$ is a tau function of the KdV hierarchy, depending only on the variables $q_\gamma=(q_{\gamma,0}, q_{\gamma,1}, q_{\gamma,2},\dots)$. We denote by $q$ the totality of the variables $q_{\gamma,n}$ for $\gamma=1,\dots,N$ and $n\geq0$.

Since each of the factors $\tau_\gamma(q_\gamma)$ satisfies the HQEs of KdV~\eqref{bil-com} with respect to the variables $q_\gamma$, the tau function $\tau(q)$ satisfies the same $N$ HQEs. 

Let us express this system of HQEs as a single equation. Let the variables $\lambda_\alpha$ be the local coordinates $\lambda-z_\alpha$ near some points $z_\alpha\in \C$, $\alpha=1,\dots,N$. 
The system of Hirota quadratic equations can be written as 
\begin{equation} \label{eq:multiplecopies}
\Res_{\lambda_\alpha}\lambda_\alpha^{2p} \, \Gamma_\alpha(\lambda_\alpha)\tau(q)\otimes \Gamma_\alpha(-\lambda_\alpha)\tau(q)   \, d\lambda=0\, ,
\end{equation}
where $p\geq 0$ and $\alpha=1,\dots,N$. Recall that the tensor product means that we are evaluating the two factors in two different sets of variables $q$ and $q'$. 
Here $\Gamma_\alpha (\lambda_\alpha )=\Gamma_{\alpha,+} ( \lambda_\alpha )\Gamma_{\alpha ,-} ( \lambda_\alpha )$, and
\begin{align}
\label{S-alpha}
\Gamma_{\alpha,-} ( \lambda_\alpha )&=
\exp\left( - {\sqrt\hbar}\sum_{n=0}^\infty b_n\lambda_\alpha ^{-2n-1}\frac{\partial}{\partial q_{\alpha ,n}}\right) \, ,
\\ \notag
\Gamma_{\alpha ,+} ( \lambda_\alpha )&=\exp \left(  \frac{1}{\sqrt\hbar}\sum_{n=0}^\infty a_n \lambda_\alpha^{2n+1}q_{\alpha,n}\right) \, ,
\end{align}
are the vertex operators of the KdV hierarchy, acting of the variables $q_\alpha$ and evaluated in the variables $\lambda_\alpha = \lambda-z_\alpha$. 



As before we can reformulate the Hirota equations for $N$ copies of KdV as regularity of the differential $1$-form in $\lambda$
\begin{equation} \label{eq:vertex-sum}
\sum_{\alpha=1}^N \left( \Gamma_\alpha(\lambda_\alpha) \otimes \Gamma_\alpha(-\lambda_\alpha) - \Gamma_\alpha(-\lambda_\alpha) \otimes \Gamma_\alpha(\lambda_\alpha) \right) (\tau\otimes\tau) \, d\lambda 
\end{equation}
where $\lambda_\alpha=\lambda-z_\alpha$. Regularity in this case means that the Laurent series expansions around each point $z_\alpha \in \C$ have no polar part. In other words, $\tau(q)$ satisfies the Hirota quadratic equations for $N$ copies of the KdV hierarchy iff~\eqref{eq:vertex-sum} is holomorphic as a function of $\lambda\in\C$.

%

Let's now obtain the Sato-Wilson and Lax equations from Hirota equations for $N$ copies of the KdV hierarchy. Let $\tau(q)$ be a solution of HQEs~\eqref{eq:multiplecopies} of the form~\eqref{asympt}.

As in the case of a single copy of the KdV hierarchy we have first to introduce a dependence of the space variable $x$ in the Hirota equations by shifting $q_{\alpha,0} \to q_{\alpha,0} +x$ for each $\alpha=1,\dots,N$.

Let
\begin{equation} \label{multau}
P_\alpha(x,q,\lambda) = \frac{\Gamma_{\alpha,-}(\lambda) \tau(x,q)}{\tau(x,q)}
\end{equation}
for $\alpha=1,\dots,N$. 

Substituting in the Hirota equations and applying the fundamental lemma we obtain the following bilinear equations in terms of pseudo-differential operators
\begin{equation} \label{SWmult}
\left( P_\alpha(x,q,\shb\partial) \Gamma_{\alpha,+}(q, \shb\partial) \cdot \hbar^p \partial^{2p} \cdot \Gamma_{\alpha,+} (q',\shb\partial)^* P_{\alpha} (x,q',\shb \partial)^* \right)_- =0
\end{equation}
for $\alpha=1,\dots,N$.

Setting $p=1$ and $q=q'$ as before, we get the constraints 
\begin{equation}
\left( P_\alpha(x,q,\shb\partial) \hbar \partial^2 P_\alpha(x,q,\shb\partial)^{-1} \right)_- = 0 .
\end{equation}
It follows that the hierarchy is described by  $N$ Lax operators given by
\begin{equation} \label{L-a}
L_\alpha = P_\alpha(x,q,\shb\partial) \hbar \partial^2 P_\alpha(x,q,\shb\partial)^{-1} = \hbar \partial^2 +2 u_\alpha(x,q).
\end{equation}

The Sato-Wilson equations are obtained by differentiating~\eqref{SWmult} w.r.t. $q_{\beta,n}$ and setting $q=q'$. Since $\Gamma_{\alpha,+}(q,\shb\partial)$ depends only on the variables  $q_\alpha$, we see that these equations decouple, as expected:
\begin{equation}
\frac{\partial P_\alpha(\shb\partial)}{\partial q_{\beta,n}} P_\alpha(\shb\partial)^{-1} = -\delta_{\alpha,\beta} \frac{a_n}{\shb} \left( P_\alpha(\shb\partial)(\shb \partial)^{2n+1} P_\alpha(\shb\partial)^{-1} \right)_- \, .
\end{equation}
Hence the dressing operator $P_\alpha$ depends only on the variables $q_\alpha$, and this in turn implies that the tau function factorizes as a product~\eqref{prodtau} of tau functions of the KdV hierarchy.

The Lax equations obviously also decouple
\begin{equation} \label{Lax-L-a}
\frac{\partial L_\alpha}{\partial q_{\beta,n}} = \delta_{\alpha,\beta} \frac{a_n}{\shb} \left[ \left( L_\alpha^{2n+1} \right)_+, L_\alpha \right] .
\end{equation}

From~\eqref{multau} we get as before the following residue formulas 
\begin{equation}
\Res_\partial P_\alpha(\shb\partial) = -\frac{\partial \log\tau}{\partial q_{\alpha,0}} 
\end{equation}
and, using Sato-Wilson equations,
\begin{equation}
\frac{\partial^2 \log\tau}{\partial q_{\alpha,0}q_{\beta,p}} = \delta_{\alpha \beta} \frac{a_n}{\shb} \Res_\partial L_\alpha^{n+\frac12}\,.
\end{equation}

Note that the factorization~\eqref{prodtau} of the tau function of multiple copies of KdV implies that the functions defined by
\begin{equation}
\Omega_{\alpha,p;\beta,q} (x,q)= \hbar \frac{\partial^2 \log\tau}{\partial q_{\alpha,p} \partial q_{\beta,q}}
\end{equation}
are non-zero only for $\alpha=\beta$.

\section{Givental-Lee infinitesimal deformations}
\label{Givental-Lee}

In this section, after recalling the basic constructions of the Givental group action on tau functions, we compute the explicit expressions of the deformations of the vertex operators and of the Hirota quadratic equations. 

\subsection{Introduction to Givental's twisted loop group action}

Let $V$ be an $N$ dimensional vector space equipped with a scalar product $\left<~,~\right>$. Fix an orthonormal basis $\{e_\alpha\}$, $\alpha=1,\dots,N$ and denote $\un = \sum_{\alpha=1}^Ne_\alpha$.

The symplectic form
\begin{equation}
\Omega(f,g) = \Res_z \left< f(-z), g(z) \right> \, dz \,,
\end{equation}
defined on the space $\cV = V((z^{-1}))$ of formal Laurent series in $z^{-1}$ with values in $V$, allows to identify $\cV$ with the cotangent bundle $T^* \cV_+$, thanks to the polarization $\cV =\cV_+ \oplus \cV_-$, where $\cV_+ = V[z]$ and $\cV_-=z^{-1} V[[z^{-1}]]$ are Lagrangian subspaces of $\cV$. 
The functions $q_{\alpha,k} = \Omega(e_\alpha (-z)^{-k-1} ,\cdot )$ and $p_{\alpha,k} = \Omega( \cdot, e_\alpha z^k )$, for $\alpha=1,\dots,N$, $k\geq0$,  define Darboux coordinates on $\cV$.

\def\cL{\mathcal{L}}
\def\GL{\mathrm{GL}}

The loop group $\cL \GL(V)$, given by $\GL(V)$-valued formal functions of $z$, acts on $\cV$. The elements of $\cL \GL(V)$ which preserve the symplectic form $\Omega$ define the {\it twisted} loop group $\cL^{(2)} \GL(V)$. The associated Lie algebra $\cL^{(2)} \End(V)$ -- given by the infinitesimal symplectic transformations of $\cV$ -- splits into two subalgebras 
\begin{equation}
\mathfrak{g}_\pm = \left\{ \mathfrak{u}(z) = \sum_{k>0} \mathfrak{u}_k z^{\pm k} , \mathfrak{u}_i \in \End (V) , \mathfrak{u}(-z)^t +\mathfrak{u}(z) =0 \right\},
\end{equation}
respectively called upper-triangular $\mathfrak{g}_+$ and lower-triangular $\mathfrak{g}_-$. 

The symplectic transformations $G(z)= e^{\mathfrak{u}(z)}\in \cL^{(2)} \GL(V)$ obtained by exponentiating elements $\mathfrak{u}(z) \in \mathfrak{g}_\pm$ define the so-called upper-triangular, $G_+$, and lower-triangular, $G_-$, subgroups of the twisted loop group. 
In the following we will typically denote by $R$ (resp. $S$) the elements of $G_+$ (resp. $G_-$).

We can associate to the elements of the twisted loop group some linear operators in the variables $q_{\alpha,p}$ by a quantization procedure which is performed in two steps. 

First, the infinitesimal symplectic trasformation of $\cV$  associated with $\mathfrak{u} \in \mathfrak{g}_\pm$ is a linear Hamiltonian vector field induced by the Hamiltonian $H_{\mathfrak{u}}(f) = \frac12\Omega(f ,\mathfrak{u} f)$, $f \in \cV$. The Hamiltonian $H_{\mathfrak{u}}$ can thus be written as a quadratic function in the Darboux variables $q_{\alpha,k}$, $p_{\alpha,k}$.

Second, $H_{\mathfrak{u}}$ can be quantized to give an operator $\hat{\mathfrak{u}}$ by the rule
\begin{equation}
q_{\alpha,k}\mapsto\frac1{\shb}{q_{\alpha,k}}, \quad
p_{\alpha,k} \mapsto \shb\frac{\partial}{\partial q_{\alpha,k}} ,
\end{equation}
with the usual ordering ambiguity fixed by 
\begin{equation}
q_{\alpha, p} p_{\beta, q} \mapsto q_{\alpha,p} \frac{\partial}{\partial q_{\beta,q}} \, .
\end{equation}

The quantization of an element $G = e^{\mathfrak{u}} \in G_\pm$ is defined by $\hat G = e^{\hat{\mathfrak{u}}}$.

Further information on Givental group action can be found in~\cite{Giv01, Lee, Lee2}.

\subsection{$R$-deformation of vertex operators}
Consider an element $\r = \r_\ell z^\ell$ of the Lie algebra $\mathfrak{g}_+$ of the twisted loop group, for fixed $\ell>0$. According to the quantization procedure described above,  the corresponding differential operator is given, up to multiplication by $(-1)^{\ell-1}$, by
\begin{align}\label{eq:first-r}
\hat \r&=\sum_{\alpha,\beta=1}^N
(\r_\ell)_{\alpha\beta}
\left(\sum_{i=0}^\infty q_{\alpha,i}\frac{\partial}{\partial  q_{\beta,\ell+i}}+
\frac{\hbar}2 \sum_{i+j=\ell-1} (-1)^{i+1}
\frac{\partial^2}{\partial q_{\alpha,i}\partial q_{\beta,j}}\right)\, ,
\end{align}
where $(\r_\ell)_{\alpha\beta} = (-1)^{\ell+1} (\r_\ell)_{\beta\alpha} =\left< e_\alpha, \r_\ell (e_\beta) \right>$ is the matrix associated to $\r_\ell \in \End(V)$.
Let $\hat R=\exp({\epsilon\hat \r})$. 

The action of the group element $\hat R$ on the vertex operators (and therefore the tangent action by the Lie algebra element $\hat \r$) is given by
\begin{equation}
\hat R \Gamma_{\alpha} \hat R^{-1} = \hat R \Gamma_{\alpha,+} \hat R^{-1} \hat R \Gamma_{\alpha,-} \hat R^{-1}\, .
\end{equation} 

A direct computation gives, up to an order $\epsilon^2$ correction, 
\begin{align}
\hat R\Gamma_{\alpha,-} ( \lambda)\hat R^{-1} & =
\Gamma_{\alpha,-} ( \lambda)
\exp\left( \epsilon{\sqrt\hbar}\sum_{\beta,n\ge 0} (\r_\ell)_{\alpha\beta}b_n\lambda^{-2n-1}\frac{\partial}{\partial q_{\beta,n+\ell}}\right) \, , \\ \notag
\hat R\Gamma_{\alpha,+} ( \lambda)\hat R^{-1} & =
\exp\left(\epsilon (\r_\ell)_{\alpha\alpha} d_\ell \lambda^{2\ell} \right)
\Gamma_{\alpha,+} ( \lambda) 
\\ \notag
&  \times \exp\left(\frac{\epsilon (-1)^{\ell-1}}{\sqrt\hbar}\sum_\beta (\r_\ell)_{\alpha\beta} 
\left(\sum_{n\ge \ell} a_n q_{\beta,n-\ell}\lambda^{2n+1}\right)\right)
\\ \notag
& \times
\exp\left(\epsilon\sqrt\hbar\sum_\beta  (\r_\ell)_{\alpha\beta}\sum_{n=0}^{\ell-1}(-1)^{n+1} a_n \frac{\partial}{\partial q_{\beta,\ell-1-n}}\lambda^{2n+1}
\right)\, ,
\end{align}
where 
\begin{equation}
d_\ell=\frac12\sum_{i+j=\ell-1}(-1)^{j+1} a_ia_j=
\begin{cases}0&\mbox{if }\ell\  \mbox{even},\\
-\frac{a_{\ell-1}}{2\ell}&\mbox{if }\ell\  \mbox{odd}.
\end{cases}
\end{equation}

To prove these identities observe that the vertex operators $\Gamma_{\alpha,\pm}(\lambda)$ are of the form $\exp \hat\f_\pm$ for certain differential operators $\hat\f_\pm$, which can be read off from~\eqref{S-alpha}. Up to $O(\epsilon^2)$ one has
\begin{align}
\hat R \Gamma_{\alpha,\pm}(\lambda) \hat R^{-1} 
&= e^{\epsilon \hat\r} e^{\hat\f_\pm} e^{-\epsilon\hat\r} 
= e^{\hat\f_\pm} ( 1 + \epsilon e^{-\hat\f_\pm} \hat\r e^{\hat\f_\pm} ) e^{-\epsilon\hat\r} \\
&= e^{\hat\f_\pm} ( 1 + \epsilon \hat\r + \epsilon\hat\a_\pm) e^{-\epsilon\hat\r}
=e^{\hat\f} e^{\epsilon\hat\a_\pm} =\Gamma_{\alpha,\pm}(\lambda) e^{\epsilon\hat\a_\pm}  \notag
\end{align}
where the operators $\hat\a_\pm$ are defined by 
\begin{equation}
e^{-\hat\f_\pm} \hat\r e^{\hat\f_\pm} = \hat\r + \hat\a_\pm \, .
\end{equation}

\begin{convention} From now on we will make all calculations up to an order $\epsilon^2$ correction. The additive term $O(\epsilon^2)$  will always be implicitly assumed.
\end{convention}

\subsection{$S$-deformation of vertex operators}
Consider an element $\s=\s_\ell z^{-\ell}$ of the Lie algebra $\mathfrak{g}_-$, for fixed $\ell>0$, and its quantization
\begin{align}
\hat\s&=
\sum_{\alpha,\beta=1}^N
(\s_\ell)_{\alpha\beta}
\left((-1)^{\ell-1} \sum_{i=0}^\infty q_{\alpha,i+\ell}\frac{\partial}{\partial  q_{\beta,i}}+
\frac1{2\hbar} \sum_{i+j=\ell-1} (-1)^{i}
 q_{\alpha,i} q_{\beta,j}\right)\, .
\end{align}
Let $\hat S=\exp({\epsilon\hat\s})$.

Again, we define the action of the group element $\hat S$ (and therefore the tangent action by the Lie algebra element $\hat\s$) as
\begin{equation}
\hat S \Gamma_{\alpha} \hat S^{-1} = \hat S \Gamma_{\alpha,+} \hat S^{-1} \hat S \Gamma_{\alpha,-} \hat S^{-1}\, .
\end{equation} 
A direct computation gives 
\begin{align}
\hat S\Gamma_{\alpha,-} & ( \lambda) \hat S^{-1}  =
\exp\left(\epsilon (\s_\ell)_{\alpha\alpha} f_\ell \lambda^{-2\ell} \right) \\ \notag
& \times\exp\left(\frac{\epsilon}{\sqrt{\hbar}}\sum_{\beta}(\s_\ell)_{\alpha\beta}\sum_{n=0}^{\ell-1} (-1)^{n} b_{n}\lambda^{-2n-1}q_{\beta,\ell-1-n}\right)
\Gamma_{\alpha,-} ( \lambda) 
\\ \notag
& \times
\exp\left(\epsilon\sqrt\hbar\sum_\beta  (\s_\ell)_{\alpha\beta}\sum_{n\geq\ell} (-1)^{\ell-1}b_n \lambda^{-2n-1} \frac{\partial}{\partial q_{\beta,n-\ell}}
\right)\\ \notag
\hat S\Gamma_{\alpha,+} & ( \lambda) \hat S^{-1}  =
\Gamma_{\alpha,+} ( \lambda) 
\\ \notag
& \times
\exp\left( \frac{\epsilon}{\sqrt\hbar}
\sum_{\beta,n\ge 0} (\s_\ell)_{\alpha\beta}a_n \lambda^{2n+1} q_{\beta,n+\ell}\right)
\, ,
\end{align}
where 
\begin{equation}
f_\ell=\frac12\sum_{i+j=\ell-1}(-1)^{i+1} b_ib_j=
\begin{cases}0&\mbox{if }\ell\  \mbox{even},\\
-\frac{b_{\ell}}{2\ell}&\mbox{if }\ell\  \mbox{odd}.
\end{cases}
\end{equation}
%
%
%

\subsection{Deformed Hirota equations}
To deform the Hirota equations~\eqref{eq:multiplecopies} we act on it with the operator $\hat R \otimes \hat R$, obtaining
\begin{align} \label{eq:mult-R-def1}
\Res_{\lambda_\alpha} \lambda_\alpha^{2p} 
&( \hat R \Gamma_{\alpha,+}(\lambda_\alpha) \hat R^{-1} )
(\hat R \Gamma_{\alpha,-}(\lambda_\alpha) \hat R^{-1}) \hat R \tau(q)  \\ \otimes
&( \hat R \Gamma_{\alpha,+}(-\lambda_\alpha) \hat R^{-1} )
(\hat R \Gamma_{\alpha,-}(-\lambda_\alpha) \hat R^{-1}) \hat R \tau(q)
 \, d\lambda_\alpha = 0 . \notag
\end{align}
We then use the formulas for the $R$-deformations of the vertex operators computed in the previous section. 

We define $R_{\alpha,+}(\lambda)$ and $R_{\alpha,-}(\lambda)$ by 
\begin{align}
\label{rplus}
R_{\alpha,+}(\lambda) : = & \exp\left(\epsilon (\r_\ell)_{\alpha\alpha} d_\ell \lambda^{2\ell} \right)
\\ \notag
&  \times \exp\left(\frac{\epsilon (-1)^{\ell-1}}{\sqrt\hbar}\sum_\beta (\r_\ell)_{\alpha\beta} 
\left(\sum_{n\ge \ell} a_n \lambda^{2n+1}q_{\beta,n-\ell}\right)\right)\, ;
\\ \label{rmin}
R_{\alpha,-}(\lambda) := & \exp\left(\epsilon\sqrt\hbar\sum_\beta  (\r_\ell)_{\alpha\beta}\sum_{n=0}^{\ell-1}(-1)^{n+1} a_n \lambda^{2n+1} \frac{\partial}{\partial q_{\beta,\ell-1-n}}
\right)
\\ \notag
& \times \exp\left( \epsilon{\sqrt\hbar}\sum_{\beta,n\ge 0} (\r_\ell)_{\alpha\beta}b_n\lambda^{-2n-1}\frac{\partial}{\partial q_{\beta,n+\ell}}\right)\, .
\end{align}
Then the $R$-deformed equation~\eqref{eq:mult-R-def1} turns into
\begin{align} \label{biltauR}
& \Res_\lambda \lambda^{2p} \left(R_{\alpha,+}(\lambda)\Gamma_{\alpha,+}(\lambda)
R_{\alpha,-}(\lambda)\Gamma_{\alpha,-}(\lambda)\hat R\tau(q) \right.
\\ \notag
& \left. \otimes R_{\alpha,+}(-\lambda)\Gamma_{\alpha,+}(-\lambda)R_{\alpha,-}(-\lambda)\Gamma_{\alpha,-}(-\lambda)\hat R\tau(q) \right) d\lambda=0\, .
\end{align}
Here $\lambda=\lambda_\alpha$ and the equation holds up to $O(\epsilon^2)$.

In a similar way, we define $S_{\alpha,+}(\lambda)$ and $S_{\alpha,-}(\lambda)$ by
\begin{align}
\label{splus}
S_{\alpha,+}(\lambda) : = & \exp\left(\epsilon (\s_\ell)_{\alpha\alpha} f_\ell \lambda^{-2\ell} \right) 
\\ \notag
& \times\exp\left(\frac{\epsilon}{\sqrt{\hbar}}\sum_{\beta}(\s_\ell)_{\alpha\beta}\sum_{n=0}^{\ell-1} (-1)^{n} b_{n}\lambda^{-2n-1}q_{\beta,\ell-1-n}\right)
\\ \notag
& \times
\exp\left( \frac{\epsilon}{\sqrt\hbar}
\sum_{\beta,n\ge 0} (\s_\ell)_{\alpha\beta}a_n\lambda^{2n+1} q_{\beta,n+\ell}\right)
\, ;
\\ \label{smin}
S_{\alpha,-}(\lambda) := & \exp\left(\epsilon\sqrt\hbar\sum_\beta  (\s_\ell)_{\alpha\beta}\sum_{n\geq\ell} (-1)^{\ell-1} b_n \lambda^{-2n-1}\frac{\partial}{\partial q_{\beta,n-\ell}}
\right)\, .
\end{align}
Then the $S$-deformed Hirota equations, obtained by acting with $\hat S \otimes \hat S$ on~\eqref{eq:multiplecopies}, turn into
\begin{align}
\label{biltauS}
& \Res_\lambda \lambda^{2p} \left(S_{\alpha,+}(\lambda)\Gamma_{\alpha,+}(\lambda)
S_{\alpha,-}(\lambda)\Gamma_{\alpha,-}(\lambda)\hat S\tau(q) \right.
\\ \notag
& \left. \otimes S_{\alpha,+}(-\lambda)\Gamma_{\alpha,+}(-\lambda)S_{\alpha,-}(-\lambda)\Gamma_{\alpha,-}(-\lambda)\hat S\tau(q) \right) d\lambda=0\, .
\end{align}
Here again $\lambda=\lambda_\alpha$ and the equation holds up to $O(\epsilon^2)$.

\subsection{Global Givental group action on vertex operators}
In the previous sections we have computed the first order approximation of the action of the Givental group on vertex operators, and considered the induced infinitesimal deformation of the Hirota equations. In fact, the action of the Givental group can be worked out globally, as follows. First we again write the vertex operator \eqref{S-alpha} as the quantization
\[
\Gamma_\alpha=e^{\hat{\f}}:=e^{\hat{\f}_-}e^{\hat{\f}_+},
\]
of the function
\begin{equation}
\label{ham}
\f(z):=\sum_{n=0}^\infty \left(-b_n\lambda_\alpha^{-2n-1}e_\alpha z^{n}-a_n\lambda_\alpha^{2n+1}e_\alpha (-z)^{-n-1}\right),
\end{equation}
and $\f_-$ resp. $\f_+$ denotes the negative resp. positive powers of $z$. According to \cite[\S 7]{Giv05}, we have
\[
Re^{\hat{\f}} R^{-1}= e^{V(\f_-^2)/2} e^{\widehat{R\f}},
\]
where the first exponent is an (explicitly computable) constant, which, as we see from \eqref{ham} only depends on positive powers of $\lambda_\alpha$. For $R=\exp(\mathfrak{r}_\ell z^\ell),~\ell\geq 0$, one computes
\[
e^{\widehat{R\f}}=\exp\left(\frac{1}{\sqrt{\hbar}}\sum_{n=0}^\infty A_{\alpha,n}q_{\alpha,n}\right)
\exp\left(-\sqrt{\hbar}\sum_{n=0}^\infty B_{\alpha,n}\frac{\partial}{\partial q_{\alpha,n}}\right),
\]
with
\begin{align*}
A_{\alpha,n}&=\sum_\beta\sum_{k=0}^\infty\frac{(-1)^{k\ell}}{k!}a_{k\ell+n}\lambda_\beta^{2(k\ell+n)+1}(\mathfrak{r}_\ell^k)_{\alpha\beta}\\
B_{\alpha,n}&=\sum_\beta\sum_{k=0}^\infty \frac{c^n_k}{k!}\lambda_\beta^{2(k\ell-n)-1}(\mathfrak{r}_\ell^k)_{\alpha\beta},\quad c^n_k=\begin{cases} b_{n-k\ell}& k\ell\leq n\\ (-1)^{k\ell-n}a_{k\ell-n-1}&k\ell>n.\end{cases} 
\end{align*}
We see that the coefficients $A_{\alpha,n}$ and $B_{\alpha,n}$ are power series containing arbitrary positive powers of $\lambda_\alpha$. This fact destroys the property of the Hirota equations \eqref{bil-com} that, after the change of variables $q_n=\xi_n+\eta_n,~q_n'=\xi_n-\eta_n$, the power series expansion in $\eta$ has coefficients that are Laurent series in $\lambda_\alpha$. Now we have arbitrary formal power series, and the Hirota equations therefore do not lead to a system of finite equations. It is this divergence that forces us to only consider the first order approximations to the Givental action, and look for a ``renormalization'' of the Hirota equations.

Similarly, the $S$-action can be worked out as (cf. \cite[\S 5]{Giv05})
\[
Se^{\hat{\f}} S^{-1}= e^{W(\f_+^2)/2} e^{\widehat{S\f}},
\]
where $e^{W(\f_+^2)/2}$ is a similar constant, which now only contains negative powers of $\lambda_\alpha$. With $S=\exp(\mathfrak{s}_\ell z^{-\ell})$, we find
\[
e^{\widehat{S\f}}=\exp\left(\frac{1}{\sqrt{\hbar}}\sum_{n=0}^\infty A_{\alpha,n}q_{\alpha,n}\right)
\exp\left(-\sqrt{\hbar}\sum_{n=0}^\infty B_{\alpha,n}\frac{\partial}{\partial q_{\alpha,n}}\right),
\]
where the coefficients are now given by
\begin{align*}
A_{\alpha,n}&=\sum_\beta\sum_{k=0}^{\infty} \frac{c^n_{k}}{k!}\lambda_\beta^{2(n-k\ell)+1}(\mathfrak{s}_\ell^k)_{\alpha\beta},\quad c^n_k=\begin{cases} (-1)^n b_{k\ell-n-1}&k\ell> n\\ (-1)^{k\ell}a_{n-k\ell}&n\leq k\ell.\end{cases} \\
B_{\alpha,n}&=\sum_\beta\sum_{k=0}^\infty \frac{b_{n+k\ell}}{k!}\lambda_\beta^{-2(n+k\ell)-1}(\mathfrak{s}_\ell^k)_{\alpha\beta}.
\end{align*}
This time $A_{\alpha,n}$ and $B_{\alpha,n}$ are bounded above in powers of $\lambda_\alpha$, and therefore, the Hirota equations for $Se^{\hat{f}} S^{-1}$ still make sense because in the expansion of $\eta$, we still have power series whose coefficients are Laurent series in $\lambda_\alpha$. Therefore, in contrast to the $R$-action, the $S$-action on the Hirota equations is indeed globally defined.

\section{Deformations of the Sato-Wilson equations}

In this section we use the formulas of the previous section to obtain deformations of the Sato-Wilson equations. 
We follow the procedure described in Section~\ref{KdV} for the KdV hierarchy and multiple copies of the KdV hierarchy.

\subsection{Equations for the deformed wave functions}
We first replace in both bilinear equations~\eqref{biltauR} and~\eqref{biltauS} the variables $q_{\alpha,0}$, for $1\le \alpha\le N$, by $q_{\alpha,0}+x$, so from now on we assume that the tau functions also depend on the variable $x$.
Next we divide equation~\eqref{biltauR} by $\hat R \tau(q)\hat R \tau(q')$ and the equation~\eqref{biltauS} by $\hat S \tau(q)\hat S \tau(q')$.
This gives the bilinear equation for the deformed wave functions.

Let $ G$ be equal to $R$ or $S$, depending on which element in the Givental group we are considering. Introduce the notation
\begin{equation} \label{PG}
P_{\alpha,G}(\lambda):=\frac{\Gamma_{\alpha,-}(\lambda)(\hat G \tau(x,q))}{\hat G\tau(x,q)}\, , \quad \alpha=1,\dots,N.
\end{equation}
We rewrite the bilinear identity~\eqref{biltauR} or~\eqref{biltauS} as follows
\begin{equation}
\label{bilV}
\Res_\lambda 
\lambda^{2p} 
V_{\alpha,G}(x,q,\lambda)e^{\frac{x\lambda}{\sqrt\hbar}}
V_{\alpha,G}(x',q',-\lambda)e^{\frac{-x'\lambda}{\sqrt\hbar}}\, d\lambda = 0,
\end{equation}
$p=0,1,2,\dots$, where
\begin{equation}
\label{V}
V_{\alpha,G}(x,q,\lambda)=
G_{\alpha,-} (\lambda)(P_{\alpha,G}(\lambda))\frac{G_{\alpha,-}(\lambda)(\hat G\tau(x,q))}{\hat G\tau(x,q)}G_{\alpha,+}(\lambda)\Gamma_{\alpha,+}(\lambda)e^{\epsilon g_{\alpha,G}(x,\lambda)}
\end{equation}
and
\begin{equation}
\label{g}
g_{\alpha,G}(x,\lambda)=
\begin{cases}
\frac{(-1)^{\ell-1}}{\sqrt\hbar}\sum_\beta \pr_{\alpha\beta}a_\ell x \lambda^{2\ell+1}
&\mbox{if }G=R,\\
\frac{(-1)^{\ell-1} }{\sqrt\hbar}\sum_\beta \ps_{\alpha\beta}b_{\ell-1} x \lambda^{-2\ell+1}&\mbox{if }G=S.
\end{cases}
\end{equation}

\subsection{Consequences of the Fundamental Lemma}
We now apply the Fundamental Lemma to the bilinear identity~\eqref{bilV}. We obtain the following system of equations for pseudo-differential operators
\begin{equation}
\label{bilV2}
\left(
V_{\alpha,G}(x,q,\sqrt\hbar\partial)\hbar^p \partial^{2p}V_{\alpha,G}(x,q',\sqrt\hbar\partial)^*\right)_-=0.
\end{equation}
In order to continue we write
\begin{equation}
\label{deformP}
V_{\alpha,G}(x,q,\sqrt\hbar\partial)=\left(P_{\alpha,G} (\sqrt\hbar\partial)+\epsilon Q_{\alpha,G}(x,q,\sqrt\hbar\partial)\right)
\Gamma_{\alpha,+}(\sqrt\hbar\partial)
\end{equation}

Clearly as before $\Gamma_{\alpha,+}(\sqrt\hbar\partial)^*=\Gamma_{\alpha,+}(\sqrt\hbar\partial)^{-1}$.

Note also that the deformed tau function $\hat G\tau$ does depend on $\epsilon$, since $\hat G$ depends on $\epsilon$.
However we will make a distintion between equations and possible tau function and $P_{\alpha,G}$ that depends on $\epsilon$.

Substitute~\eqref{deformP} into~\eqref{bilV2}; this gives for $p=0,1,2,\dots$ and $q=q'$
\begin{equation} \label{bilV3}
\left(
\left(P_{\alpha,G} +\epsilon Q_{\alpha,G}\right)
\hbar^p \partial^{2p}
\left(P_{\alpha,G}^*+\epsilon Q_{\alpha,G}^*\right)\right)_-=0\, . 
\end{equation}

Observe first that, by definition, $\left(P_{\alpha,G}(\sqrt\hbar\partial)P_{\alpha,G}(\sqrt\hbar\partial)^*\right)_+=1$. 
Thus~\eqref{bilV3} for $p=0$ gives 
\begin{equation}
\epsilon P_{\alpha,G}^* = \epsilon P_{\alpha,G}^{-1}
\end{equation}
(by this notation we mean that the identity holds at the zeroth order $\epsilon$). Taking into account also the first order in $\epsilon$ we get
\begin{align}
\label{bilV4}
P_{\alpha,G}^*+\epsilon Q_{\alpha,G}^*&=P_{\alpha,G}^{-1} +\epsilon Q_{\alpha,G}^*-\epsilon
P_{\alpha,G}^{-1}\left(Q_{\alpha,G}P_{\alpha,G}^{-1}+P_{\alpha,G}Q_{\alpha,G}^*\right)_-
\\ \notag
&=P_{\alpha,G}^{-1}\left(1-\epsilon Q_{\alpha,G} P_{\alpha,G}^{-1}+\epsilon\left(Q_{\alpha,G}P_{\alpha,G}^{-1}+P_{\alpha,G}Q_{\alpha,G}^*\right)_+ \right)
\, .
\end{align}

\subsection{The Lax operator}
Define the deformed $\alpha$-th Lax operator as 
\begin{equation} \label{LG}
L_{\alpha,G}:=P_{\alpha,G}\hbar\partial^2P_{\alpha,G}^{-1}=L_\alpha+\left(L_{\alpha,G}\right)_-\, ,
\end{equation}
where $L_\alpha$ denotes its differential part, which is necessarily of the form 
\begin{equation}
L_\alpha=\hbar\partial^2 +2u_\alpha(x,q,\epsilon) \, .
\end{equation}
We stress here that $L_\alpha$, and $u_\alpha$, do depend on the deformation parameter $\epsilon$, since they are computed from the deformed tau function $\hat G \tau$. 
As we already know, in the undeformed case the negative part of the Lax operator vanishes, hence here it must be at least of order $\epsilon$. In other words $\epsilon L_{\alpha,G}=\epsilon L_\alpha$. 

Now let's compute how the constraints on the Lax operator look like at the first order in $\epsilon$. 
Equation (\ref{bilV3}) for $p=1$ gives 
\begin{equation}
\label{1}
\left(P_{\alpha,G}\hbar\partial^2P_{\alpha,G}^*+\epsilon Q_{\alpha,G}\hbar\partial^2P_{\alpha,G}^*+\epsilon P_{\alpha,G}\hbar\partial^2Q_{\alpha,G}^*\right)_-=0\, .
\end{equation}
Substituting here~\eqref{bilV4} we obtain
%
\begin{equation}
\label{3}
\left(L_{\alpha,G}-\epsilon [ L_{\alpha,G} , Q_{\alpha,G} P_{\alpha,G}^{-1} ] + \epsilon L_{\alpha,G} (Q_{\alpha,G} P_{\alpha,G}^{-1} + P_{\alpha,G} Q_{\alpha,G}^*)_+ \right)_-=0\, .
\end{equation}

Thus
\begin{equation}
\label{newL}
L_{\alpha,G}=L_\alpha+\epsilon [L_{\alpha},Q_{\alpha,G} P_{\alpha,G}^{-1}]_-\, .
\end{equation}

\subsection{The deformed Sato-Wilson equations}

In order to obtain the deformed Sato-Wilson equations we differentiate the first component of  the Hirota bilinear identity~\eqref{bilV2} w.r.t. $q_{\beta,n}$ and set $p=0$, $q=q'$,
\begin{align}
\label{4}
0=&\left(
\left(
\frac{\partial P_{\alpha,G}}{\partial q_{\beta,n}} +\epsilon \frac{\partial Q_{\alpha,G}}{\partial q_{\beta,n}}
+\delta_{\alpha\beta}\frac{a_n}{\sqrt\hbar}
\left( P_{\alpha,G}+\epsilon Q_{\alpha,G} \right)
\left(\sqrt\hbar\partial\right)^{2n+1}
\right)  
\right. \\ \notag 
& \left.\phantom{\frac{\partial P_{\alpha,G}}{\partial q_{\beta,n}}} \times \left(P_{\alpha,G}^*+\epsilon Q_{\alpha,G}^*\right)
\right)_-
\\ \notag
=&
\left(
\left(
\frac{\partial P_{\alpha,G}}{\partial q_{\beta,n}} +\epsilon \frac{\partial Q_{\alpha,G}}{\partial q_{\beta,n}}
+\delta_{\alpha\beta}\frac{a_n}{\sqrt\hbar}
\left(
P_{\alpha,G}+\epsilon Q_{\alpha,G}
\right)\left(\sqrt\hbar\partial\right)^{2n+1}
\right)
P_{\alpha,G}^{-1}
\right.
\\ \notag
& \left.\phantom{\frac{\partial P_{\alpha,G}}{\partial q_{\beta,n}}}  
\times
\left(1-\epsilon\left(Q_{\alpha,G}P_{\alpha,G}^{-1}-\left(Q_{\alpha,G}P_{\alpha,G}^{-1}+P_{\alpha,G}Q_{\alpha,G}^*\right)_+\right)\right)
\right)_-
\, .
\end{align}
Thus we see that at the leading order we have the usual Sato-Wilson equations for $N$ copies of the KdV hierarchy
\begin{equation}
\label{5}
\frac{\partial P_{\alpha,G}}{\partial q_{\beta,n}} P_{\alpha,G}^{-1}
+\delta_{\alpha\beta}\frac{a_n}{\sqrt\hbar}\left(
L_{\alpha,G}^{n+\frac12}
\right)_-=O(\epsilon) \, .
\end{equation}
Using this we can rewrite Equation~\eqref{4} as
\begin{align}
\label{newSato}
&\frac{\partial P_{\alpha,G}}{\partial q_{\beta,n}} P_{\alpha,G}^{-1}
+\delta_{\alpha\beta}\frac{a_n}{\sqrt\hbar}\left(
L_{\alpha,G}^{n+\frac12}
\right)_-
\\ \notag
&+\epsilon
\left(
\frac{\partial Q_{\alpha,G}P_{\alpha,G}^{-1}}{\partial q_{\beta,n}}-\delta_{\alpha\beta}\frac{a_n}{\sqrt\hbar}\left[
\left(L_\alpha^{n+\frac12}\right)_+,
Q_{\alpha,G}P_{\alpha,G}^{-1}\right]
\right)_-=0\, ,
\end{align}
which is a deformation of the Sato-Wilson equations.

In principle one can use this equation (\ref{newSato}) to obtain  an expression, a Lax type equation, for $\frac{\partial L_{\alpha,G}}{\partial q_{\beta,n}} $.

\subsection{An alternative form of the deformed Sato-Wilson equations}
Later on it will be convenient to use a different form of the deformed Sato-Wilson equations.
We define 
\begin{equation}
\label{newLL}
\tilde L_{\alpha,G}:=L_\alpha-\epsilon [L_\alpha,Q_{\alpha,G}P_{\alpha,G}^{-1}]_+ = L_{\alpha,G}- \epsilon [L_\alpha,Q_{\alpha,G}P_{\alpha,G}^{-1}]\, .
\end{equation}
Since the commutator on the right-hand side is not projected, it is easy to compute the square root of $\tilde L_{\alpha,G}$, and consequently
\begin{equation}
\left(L_{\alpha,G}^{n+\frac12}\right)_- = \left(\tilde L_{\alpha,G}^{n+\frac12} + \epsilon [L_\alpha^{n+\frac{1}{2}},Q_{\alpha,G}P_{\alpha,G}^{-1}] \right)_-\, .
\end{equation}
Substituting in the deformed Sato-Wilson equations~\eqref{newSato}, we rewrite them as
\begin{align}
\label{newSato2}
&\frac{\partial P_{\alpha,G}}{\partial q_{\alpha,p}} P_{\alpha,G}^{-1}
+\frac{a_p}{\sqrt\hbar}\left(
\tilde L_{\alpha,G}^{p+\frac12}
\right)_-
\\ \notag
&+\epsilon
\left(
\frac{\partial Q_{\alpha,G}P_{\alpha,G}^{-1}}{\partial q_{\alpha,p}}+\frac{a_p}{\sqrt\hbar}\left[
\left(L_\alpha^{p+\frac12}\right)_-,
Q_{\alpha,G}P_{\alpha,G}^{-1}\right]
\right)_-=0
\end{align}
in the case $\alpha=\beta$. Note that the main advantage of this formula is the different sign of the projector appearing in the commutator.

If $\alpha\not=\beta$, we still have
\begin{equation}
\label{newSato3}
\frac{\partial P_{\alpha,G}}{\partial q_{\beta,p}} P_{\alpha,G}^{-1}
+\epsilon
\frac{\partial Q_{\alpha,G}P_{\alpha,G}^{-1}}{\partial q_{\beta,p}}=0\, .
\end{equation}

\subsection{Explicit formulas for $Q_{\alpha,G}$ and $Q_{\alpha,G}P_{\alpha,G}^{-1}$}
Now it is straightforward to check that
\begin{align}
\label{QR}
Q_{\alpha,R} & = \pr_{\alpha\alpha} d_\ell P_{\alpha,R}(\sqrt\hbar\partial)(\sqrt\hbar\partial)^{2\ell}
\\ \notag
&+{\sqrt\hbar}{\pr_{\alpha\alpha}} \sum_{n=0}^{\ell-1}(-1)^{n+1} a_n \frac{\partial P_{\alpha,R}}{\partial q_{\alpha,\ell-1-n}}(\sqrt\hbar\partial)^{2n+1}
\\ \notag
&+{\sqrt\hbar}{\pr_{\alpha\alpha}} \sum_{n\ge 0} b_n \frac{\partial P_{\alpha,R}}{\partial q_{\alpha,n+\ell}}(\sqrt\hbar\partial)^{-2n-1}
\\ \notag
&+\frac{(-1)^{\ell-1}}{\sqrt\hbar} \sum_\beta \pr_{\alpha\beta} a_\ell x P_{\alpha,R}(\sqrt\hbar\partial)(\sqrt\hbar\partial)^{2\ell+1}
\\ \notag
&+ \frac{(-1)^{\ell-1}}{\sqrt\hbar} \sum_\beta \pr_{\alpha\beta} \sum_{n\geq \ell} a_n q_{\beta,n-\ell}
P_{\alpha,R}(\sqrt\hbar\partial)(\sqrt\hbar\partial)^{2n+1}
\\ \notag
&+{\sqrt\hbar} \sum_\beta \pr_{\alpha\beta}
\sum_{n=0}^{\ell-1}(-1)^{n+1} a_n \frac{\partial \log \tau}{\partial  q_{\beta,\ell-1-n}}P_{\alpha,R}(\sqrt\hbar\partial)(\sqrt\hbar\partial)^{2n+1}
\\ \notag
& +{\sqrt\hbar} \sum_\beta \pr_{\alpha\beta}
\sum_{n\geq 0} b_n \frac{\partial\log\tau}{\partial q_{\beta,\ell+n}}P_{\alpha,R}(\sqrt\hbar\partial)(\sqrt\hbar\partial)^{-2n-1} .
\end{align}
This formula is obtained by identifying in~\eqref{V} the $O(\epsilon)$ contributions after substitution of~\eqref{rplus},\eqref{rmin}, and comparing with the definition~\eqref{deformP} of $Q_{\alpha,R}$. One has to carefully take into account the correct ordering of the operators by placing powers of $\lambda$ on the right before substituting $\lambda$ with $\shb\partial$. 

Note that $Q_{\alpha,G}$ enters the deformed Sato-Wilson equations in terms which are of order at least $\epsilon$. For this reason here we are proving these identities only up to $O(\epsilon)$.

Similarly we get
\begin{align}
\label{QS}
Q_{\alpha,S}& =
\ps_{\alpha\alpha} f_\ell P_{\alpha,S}(\sqrt\hbar\partial) (\sqrt\hbar\partial)^{-2\ell}
\\ \notag &
+{\sqrt\hbar} \ps_{\alpha\alpha}
\sum_{n\ge \ell} b_n\frac{\partial P_{\alpha,S}}{\partial q_{\alpha,n-\ell}}(\sqrt\hbar\partial)^{-2n-1}
\\ \notag &
+\frac{1}{\sqrt\hbar}\sum_\beta
\ps_{\alpha\beta} (-1)^{\ell-1} b_{\ell-1} x  P_{\alpha,S}(\sqrt\hbar\partial)(\sqrt\hbar\partial)^{-2\ell+1} 
\\ \notag &
+\frac{1}{\sqrt\hbar}\sum_\beta
\ps_{\alpha\beta}  
\sum_{n=0}^{\ell-1}(-1)^{n} b_n q_{\beta,\ell-1-n} P_{\alpha,S}(\sqrt\hbar\partial)(\sqrt\hbar\partial)^{-2n-1}
\\ \notag &
+\frac{1}{\sqrt\hbar}\sum_\beta
\ps_{\alpha\beta}  
\sum_{n\ge 0} a_n q_{\beta,\ell+n} P_{\alpha,S}(\sqrt\hbar\partial)(\sqrt\hbar\partial)^{2n+1}
\\ \notag &
+ {\sqrt\hbar} \sum_\beta \ps_{\alpha\beta} 
(-1)^{\ell-1}
\sum_{n\ge \ell} b_n \frac{\partial \log \tau}{\partial q_{\beta,n-\ell}} P_{\alpha,S}(\sqrt\hbar\partial)(\sqrt\hbar\partial)^{-2n-1}\, .
\end{align}
Thus, multiplying on the right by $P_{\alpha,G}^{-1}$, we get
\begin{align}
\label{QRP}
Q_{\alpha,R}P_{\alpha,R}^{-1} & = \pr_{\alpha\alpha} d_\ell L_{\alpha,R}^{\ell}
\\ \notag
&-{\pr_{\alpha\alpha}} \sum_{n=0}^{\ell-1}(-1)^{n+1} a_n a_{\ell-1-n} \left(L_{\alpha,R}^{\ell-n-\frac{1}{2}}\right)_- L_{\alpha,R}^{n+\frac{1}{2}}
\\ \notag
&-{\pr_{\alpha\alpha}} \sum_{n\ge 0} b_n a_{n+\ell} \left(L_{\alpha,R}^{n+\ell+\frac{1}{2}}\right)_- L_{\alpha,R}^{-n-\frac{1}{2}}
\\ \notag
&+\frac{(-1)^{\ell-1}}{\sqrt\hbar} \sum_\beta \pr_{\alpha\beta} a_\ell x L_{\alpha,R}^{\ell+\frac12}
\\ \notag
&+ \frac{(-1)^{\ell-1}}{\sqrt\hbar} \sum_\beta \pr_{\alpha\beta} \sum_{n\geq \ell} a_n q_{\beta,n-\ell} L_{\alpha,R}^{n+\frac12}
\\ \notag
&+{\sqrt\hbar} \sum_\beta \pr_{\alpha\beta}
\sum_{n=0}^{\ell-1}(-1)^{n+1} a_n \frac{\partial \log \tau}{\partial  q_{\beta,\ell-1-n}}L_{\alpha,R}^{n+\frac12}
\\ \notag
& +{\sqrt\hbar} \sum_\beta \pr_{\alpha\beta}
\sum_{n\geq 0} b_n \frac{\partial\log\tau}{\partial q_{\beta,\ell+n}}L_{\alpha,R}^{-n-\frac12}
\end{align}
and
\begin{align}
\label{QSP}
Q_{\alpha,S}P_{\alpha,S}^{-1}& =
\ps_{\alpha\alpha} f_\ell L_{\alpha,S}^{-\ell}
\\ \notag &
- \ps_{\alpha\alpha}
\sum_{n\ge \ell} b_n a_{n-\ell} \left(L_{\alpha,S}^{n-\ell+\frac12}\right)_- L_{\alpha,S}^{-n-\frac12}
\\ \notag &
+\frac{1}{\sqrt\hbar}\sum_\beta
\ps_{\alpha\beta} (-1)^{\ell-1} b_{\ell-1} x  L_{\alpha,S}^{-\ell+\frac12} 
\\ \notag &
+\frac{1}{\sqrt\hbar}\sum_\beta
\ps_{\alpha\beta}  
\sum_{n=0}^{\ell-1}(-1)^{n} b_n q_{\beta,\ell-1-n} L_{\alpha,S}^{-n-\frac12}
\\ \notag &
+\frac{1}{\sqrt\hbar}\sum_\beta
\ps_{\alpha\beta}  
\sum_{n\ge 0} a_n q_{\beta,\ell+n} L_{\alpha,S}^{n+\frac12}
\\ \notag &
+ {\sqrt\hbar} \sum_\beta \ps_{\alpha\beta} (-1)^{\ell-1}
\sum_{n\ge \ell} b_n \frac{\partial \log \tau}{\partial q_{\beta,n-\ell}} L_{\alpha,S}^{-n-\frac12}\, .
\end{align}
Since we are computing these identities up to terms $O(\epsilon)$, we may substitute everywhere $L_{\alpha,G}$ with $L_\alpha$.

\subsection{Explicit computation of $S$-deformations}

Let us introduce the deformed $\Omega$ functions
\begin{equation}
\hat G \Omega_{\alpha,p;\beta,q} := \hbar \frac{\partial^2 \log \hat G \tau}{\partial q_{\alpha,p} \partial q_{\beta,q}} \,.
\end{equation}
As in the undeformed case we get from~\eqref{PG} that
\begin{equation} \label{Gome}
\hat G \Omega_{\alpha,0;\beta,p} = - \Res_\partial \frac{\partial P_{\alpha,G}}{\partial q_{\beta,q}} P_{\alpha,G}^{-1}\,.
\end{equation}

Now we proceed to substitute in this equation the deformed Sato-Wilson equations and the explicit formulas for $Q_{\alpha,G} P_{\alpha,G}^{-1}$ obtained above, hence obtaining deformation formulas to be compared to those derived in the Hamiltonian approach in~\cite{BurPosSha1}. 

Let us begin with the $S$-deformations. 
In the following it is more convenient, from a notational point of view, to consider a general element $\s =\sum_{\ell\geq1} \s_\ell z^{-\ell}$ of the lower triangular Lie algebra (rather than fixing $\ell$).
As before $\hat S = e^{\epsilon\hat\s}$.

In the case $\alpha\not=\beta$ we have from Equation~\eqref{newSato3}
\begin{align}\label{eq:ResNewSatoab}
& \frac{\hat{S}\Omega_{\alpha,0;\beta,p}}{\hbar} = \epsilon \Res_\partial \frac{\d Q_{\alpha,S}P_{\alpha,S}^{-1}}{\d q_{\beta,p}} 
\\ \notag
& = \frac{\epsilon}{\sqrt{\hbar}} \left( (\s_{p+1})_{\alpha\beta} \Res_\d L_\alpha^{-\frac12} 
+ \sum_{\ell=1}^{p} (\s_{\ell})_{\alpha\beta} a_{p-\ell} \Res_\d L_\alpha^{p-\ell+\frac12} \right)
\\ \notag
& = \frac{\epsilon}{\hbar} \left( (\s_{p+1})_{\alpha\beta}+\sum_{\ell=1}^{p} (\s_{\ell})_{\alpha\beta}  \Omega_{\alpha,0;\alpha,p-\ell} \right)\, .
\end{align}
Note that this espression has no constant term in $\epsilon$, since in the case of several copies of KdV the undeformed $\Omega_{\alpha,0;\beta,p}$, $\alpha\not=\beta$ is equal to zero.

Let $\alpha=\beta$. Then Equation~\eqref{newSato2} implies that
\begin{align} \label{eq:ResDefS}
\frac{\hat{S}\Omega_{\alpha,0;\alpha,p}}{\hbar} = & \frac{a_p}{\sqrt\hbar} \Res_\d \tilde L_{\alpha,S}^{p+\frac12} \\ \notag &
+ \epsilon \Res_\partial \left( \frac{\d Q_{\alpha,S}P_{\alpha,S}^{-1}}{\d q_{\alpha,p}} +\frac{a_p}{\sqrt\hbar} 
\left[\left(L_\alpha^{p+\frac12}\right)_-, Q_{\alpha,S}P_{\alpha,S}^{-1}\right]\right) 
\end{align}
This expression requires some further computation.

\begin{lemma} \label{lem:ResLtilde} We have: $a_p\sqrt\hbar\Res_\d \tilde L_{\alpha,S}^{p+\frac12} = \Omega_{\alpha,0;\alpha,p} - \epsilon(\s_1)_{\alpha\un} \Omega_{\alpha,0;\alpha,p-1}$.
\end{lemma}
\begin{proof}
The square root of $\tilde L_{\alpha,S}$ is of the form
\begin{equation}
\tilde L_{\alpha,S}^{\frac12} = L_\alpha^{\frac12} +\epsilon Y
\end{equation}
where $Y$ is a pseudo-differential operator that solves the equation
\begin{equation}
YL_\alpha^{\frac12}+L_\alpha^{\frac12}Y=-[L,Q_{\alpha,S}P_{\alpha,S}^{-1}]_+\, .
\end{equation}
It is clear from~\eqref{QSP} that 
\begin{equation}
-[L_\alpha,Q_{\alpha,S}P_{\alpha,S}^{-1}]_+
= 
-\frac{1}{\sqrt\hbar} \left(
(\s_1)_{\alpha\un} [L_\alpha,x]  L_\alpha^{-\frac12} \right)_+
=-2(\s_1)_{\alpha\un} \,,
\end{equation}
therefore, $Y=- (\s_1)_{\alpha\un} L_\alpha^{-\frac12}$ .
Hence we have
\begin{equation}
\tilde L_{\alpha,S}^{p+\frac12} = L_\alpha^{p+\frac12} - \epsilon (2p+1) (s_1)_{\alpha,\un} L_\alpha^{p-\frac12} \,.
\end{equation}�
Taking the residue of this expression and recalling that 
\begin{equation}
\frac{\Omega_{\alpha,0;\alpha,p}}\hbar = \frac{a_p}{\shb} \Res_\partial L_\alpha^{p+\frac12}
\end{equation}
and that $(2p+1)a_p=a_{p-1}$, the statement of the Lemma follows.
%
\end{proof}

The second summand on the right hand side of Equation~\eqref{eq:ResDefS} can be computed directly. The Lax equations imply that at the first order in $\epsilon$ the operator
\begin{equation}
X\mapsto
\frac{\d X}{\d q_{\alpha,p}} +\frac{a_p}{\sqrt\hbar} 
\left[\left(L_\alpha^{p+\frac12}\right)_-,X\right]
\end{equation} 
vanishes when applied to any power of $L_\alpha$. One can easily check that only a few summands in Equation~\eqref{QSP} can contribute to the residue, and a direct computation shows that 
\begin{align}    \label{eq:LasSDef}
& \Res_\partial \left( \frac{\d Q_{\alpha,S}P_{\alpha,S}^{-1}}{\d q_{\alpha,p}} +\frac{a_p}{\sqrt\hbar} 
\left[\left(L_\alpha^{p+\frac12}\right)_-, Q_{\alpha,S}P_{\alpha,S}^{-1}\right]\right)
\\ \notag &
 = \frac{\epsilon}{\sqrt{\hbar}} \left( (\s_{p+1})_{\alpha\alpha} \Res_\d L_\alpha^{-\frac12} 
+ \sum_{\ell=1}^{p} (\s_{\ell})_{\alpha\alpha} a_{p-\ell} \Res_\d L_\alpha^{p-\ell+\frac12} \right)
\\ \notag
& = \frac{\epsilon}{\hbar} \left( (\s_{p+1})_{\alpha\alpha}+\sum_{\ell=1}^{p} (\s_{\ell})_{\alpha\alpha} \Omega_{\alpha,0;\alpha,p-\ell} \right) \,.
\end{align}

Summarizing, for $\alpha =\beta$ we have
\begin{align}
\hat S \Omega_{\alpha,0;\alpha,p} = \Omega_{\alpha,0;\alpha,p} + \epsilon \left( (\s_{p+1})_{\alpha\alpha} + \sum_{\ell=1}^p (\s_\ell)_{\alpha\alpha} \Omega_{\alpha,0;\alpha,p-\ell} - (\s_1)_{\alpha\un} \Omega_{\alpha,0;\alpha,p-1} \right) \,.
\end{align}

\subsection{Explicit computation of $R$-deformations}
Let us now consider the deformation formulas obtained by substituting the Sato-Wilson equations and the explicit formula for $Q_{\alpha,R}P_{\alpha,R}^{-1}$ in equation~\eqref{Gome}. Here we still consider $\r = \r_\ell z^\ell$ for $\ell$ fixed.

First we consider the case $\alpha\not=\beta$. Equation~\eqref{newSato3} implies that
\begin{align}\label{eq:R-ResNewSatoab}
& \frac{\hat{R}\Omega_{\alpha,0;\beta,p}}{\hbar}  = \epsilon \Res_\partial \frac{\d Q_{\alpha,R}P_{\alpha,R}^{-1}}{\d q_{\beta,p}} 
\\ \notag
& = \frac{\epsilon(\r_\ell)_{\alpha\beta}}{\sqrt\hbar} \Res_\partial
\left( (-1)^{\ell-1} a_{p+\ell} L_\alpha^{p+\ell+\frac12} \phantom{\sum_{i=0}^{\ell-1}}\right.
\\ \notag 
& \left.
+\sum_{i=0}^{\ell-1}(-1)^{i+1} a_i \Omega_{\beta,\ell-1-i;\beta,p} L_\alpha^{i+\frac12}
+\Omega_{\beta,\ell;\beta,p} L_\alpha^{-\frac12} \right)
\\ \notag
& = \frac{\epsilon(\r_\ell)_{\alpha\beta}}{\hbar}
\left((-1)^{\ell-1} \Omega_{\alpha,0;\alpha,p+\ell} \phantom{\sum_{i=0}^{\ell-1}}\right.
\\ \notag & + \left.
\sum_{i=0}^{\ell-1}(-1)^{i+1} \Omega_{\alpha,0;\alpha,i}\Omega_{\beta,\ell-1-i;\beta,p}
+ \Omega_{\beta,\ell;\beta,p}\right) \,.
\end{align}

In the case $\alpha=\beta$ the residue of the Equation~\eqref{newSato3} turns into
\begin{align}\label{eq:R-aa}
& \frac{\hat{R}\Omega_{\alpha,0;\alpha,p}}{\hbar}  = \frac{a_p}{\sqrt\hbar} \Res_\d \left( \tilde L_{\alpha,R} \right)^{p+\frac12} 
\\ \notag
& + \epsilon \Res_\d \left(
-{(\r_\ell)_{\alpha\alpha}} \sum_{i=0}^{\ell-1}(-1)^{i+1} a_i a_{p} 
\frac{\d \left(L_\alpha^{p+\frac{1}{2}}\right)_-}{\d q_{\alpha,\ell-i-1}} L_\alpha^{i+\frac{1}{2}} 
\right.
\\ \notag
&+\frac{(-1)^{\ell-1}}{\hbar} (\r_\ell)_{\alpha\un} a_\ell a_p \left[\left(L_\alpha^{p+\frac12}\right)_-, x\right] L_\alpha^{\ell+\frac12}
\\ \notag
&+ \frac{1}{\sqrt\hbar} (\r_\ell)_{\alpha\alpha} a_{p+\ell} L_\alpha^{p+\ell+\frac12}
\\ \notag
&+\frac1{\sqrt\hbar} (\r_\ell)_{\alpha\alpha}
\sum_{i=0}^{\ell-1}(-1)^{i+1} a_i \Omega_{\alpha,\ell-1-i;\alpha,p}L_\alpha^{i+\frac12}
\\ \notag
&+\sum_\beta (\r_\ell)_{\alpha\beta}
\sum_{i=0}^{\ell-1}(-1)^{i+1} a_i a_p \left[ \left(L_\alpha^{p+\frac12}\right)_-, \frac{\d\log\tau}{\d q_{\beta,\ell-1-i}} \right] L_\alpha^{i+\frac12}
\\ \notag
& \left. +\frac1{\sqrt\hbar} (\r_\ell)_{\alpha\alpha}
\Omega_{\alpha,\ell;\alpha,p}L_\alpha^{-\frac12} \right)\, .
\end{align}
This is a straightforward computation, one has just to use the zero-curvature equations~\eqref{ZS}. 

As in Lemma~\ref{lem:ResLtilde}, we need to compute the first residue on the right-hand side of~\eqref{eq:R-aa}. The square root of $\tilde L_{\alpha,R}$ is of the form
$
\tilde L_{\alpha,R}^{\frac12} = L_{\alpha}^{\frac12} + \epsilon Y
$
where $Y$ is a pseudo-differential operator that solves the equation
\begin{equation}\label{eq:Y}
YL_\alpha^{\frac12}+L_\alpha^{\frac12}Y=-[L_\alpha,Q_{\alpha,R}P_{\alpha,R}^{-1}]_+ .
\end{equation}
Then, we have 
\begin{equation} \label{YY125}
a_p\sqrt\hbar\Res_\d \tilde L_{\alpha,R}^{p+\frac12}=\Omega_{\alpha,0;\alpha,p}+\epsilon a_p\sqrt\hbar\Res_\d X
\end{equation}
where $X= YL_\alpha^{p}+L_\alpha^{\frac12}YL_\alpha^{p-\frac12}+\cdots+L_\alpha^pY$.

Explicitly, equation~\eqref{eq:Y} becomes
\begin{align}
YL_\alpha^{\frac12}+L_\alpha^{\frac12}Y
 &= \left(
{(\r_\ell)_{\alpha\alpha}} \sum_{i=0}^{\ell-1}(-1)^{i+1} a_i a_{\ell-1-i} \left[ L_\alpha, \left(L_\alpha^{\ell-i-\frac{1}{2}}\right)_-\right] L_\alpha^{i+\frac{1}{2}}
\right. 
\\ \notag
&-\frac{(-1)^{\ell-1}}{\sqrt\hbar} (\r_\ell)_{\alpha\un} a_\ell [L_\alpha,x] L_\alpha^{\ell+\frac12}
\\ \notag
&-{\sqrt\hbar} \sum_\beta (\r_\ell)_{\alpha\beta}
\sum_{i=0}^{\ell-1}(-1)^{i+1} a_i \left[ L_\alpha, \frac{\partial \log \tau}{\partial  q_{\beta,\ell-1-i}}\right] L_\alpha^{i+\frac12}
\\ \notag 
& \left. -{\sqrt\hbar} \sum_\beta (\r_\ell)_{\alpha\beta}
\left[ L_\alpha, \frac{\partial\log\tau}{\partial q_{\beta,\ell}}\right] L_\alpha^{-\frac12}
\right)_+ \, .
\end{align}
Surprisingly enough, this equation, whose solution we could reasonably expect to have only implicitly, has an explicit solution. We give it below, in Lemma~\ref{lem:solY}.

%

\section{Comparison with deformations in Hamiltonian form}

In this section we recall the deformation formulas obtained in~\cite{BurPosSha1} in the Hamiltonian approach and compare them with the formulas we have just obtained via deformation of vertex operators.

\subsection{The $R$-deformations}
We begin with the deformation formula for the $\Omega_{\alpha,p;\beta,q}$ obtained in~\cite[Thm. 7]{BurPosSha1}.

\begin{align} \label{eq:def-Omega}
& \widehat{\r_\ell z^\ell}[u]. \Omega _{\alpha,p;\beta,q} = 
 (\r_\ell)_\alpha^\mu \Omega_{\mu,p+\ell;\beta,q} + \Omega_{\alpha,p;\mu,q+\ell}(\r_\ell)_\beta^\mu \\ \notag
& \phantom{ =\ } +\sum_{i=0}^{\ell-1} (-1)^{i+1} \Omega_{\alpha,p;\mu,i} (\r_\ell)^{\mu\nu} \Omega_{\nu,\ell-1-i;\beta,q} \\ \notag
& \phantom{ =\ } - \sum_{\gamma,n} \frac{\d \Omega _{\alpha,p;\beta,q}}{\d u_{\gamma,n}} \left(  
(\r_\ell)^\mu_\gamma \d_x^n \Omega_{\mu,\ell;\un,0} 
+ (n+1)\d_x^n \Omega _{\gamma,0;\mu,\ell} (\r_\ell)^\mu_\un \phantom{\sum_{i=1}^{\ell-1}}
\right. \\ \notag
& \phantom{ = -\ }
+\sum_{i=0}^{\ell-1}\sum_{k=0}^{n-1} \binom{n}{k} (-1)^{i+1} \d_x^{k+1}\Omega_{\gamma,0;\mu,i} (\r_\ell)^{\mu\nu} \d_x^{n-k-1}\Omega_{\nu,\ell-1-i;\un,0} \\
& \phantom{ = -\ } \notag
\left.
+\sum_{i=0}^{\ell-1} (-1)^{i+1} \d_x^n \left(\Omega_{\gamma,0;\mu,i} (\r_\ell)^{\mu\nu} \Omega_{\nu,\ell-1-i;\un,0} \right)
\right) \\ \notag
& 
+\frac{\hbar}{2} \sum_{
\begin{smallmatrix}
\gamma, n\\
\zeta, m
\end{smallmatrix}} \frac{\d^2 \Omega _{\alpha,p;\beta,q}}{\d u_{\gamma,n}\d u_{\zeta,m}}
\sum_{i=0}^{\ell-1}(-1)^{i+1}
\d_x^{n+1} \Omega_{\gamma,0;\mu,i} (\r_\ell)^{\mu\nu} \d_x^{m+1} \Omega_{\nu,\ell-1-i;\zeta,0}.
\end{align}
In this formula, a subscript $\un$ means summing over all indices, e.g., $\Omega_{\mu,\ell;\un,0}=\sum_\nu \Omega_{\mu,\ell;\nu,0}$,  and the following sign convention of \cite{FSZ} is used for raising and lowering indices
\begin{equation} \label{raising}
(\r_\ell)^{\beta\alpha} = (\r_\ell)^\alpha_\beta = (\r_\ell)_{\alpha\beta}\,.
\end{equation}
Recall that the symmetry properties of $\r_\ell$ are simply  
\begin{equation}
(\r_\ell)_{\alpha \beta} = (-1)^{\ell+1} (\r_\ell)_{\beta\alpha}\,.
\end{equation}
Since we are considering deformations of the tau function of multiple copies the KdV hierarchy, we have that 
$\Omega_{\alpha,p;\beta,q}=0$ if $\alpha\not=\beta$. 

Moreover it will be sufficient to consider only the case $p=0$, since the deformations of $\Omega_{\alpha,0;\beta,q}$ completely determine the Hamiltonian structure of the deformed hierarchy.

Taking into account these restrictions, the general formula~\eqref{eq:def-Omega}, in the case $\alpha\not=\beta$, reduces to 
\begin{align} \label{eq:def-Omega-a-b}
\widehat{\r_\ell z^\ell}[u]. \Omega _{\alpha,0;\beta,p} & = 
(-1)^{\ell-1} (\r_\ell)_{\alpha\beta}\left( \Omega_{\beta,\ell;\beta,p} + (-1)^{\ell-1} \Omega_{\alpha,0;\alpha,p+\ell} \phantom{\sum_{i=0}^{\ell-1}} \right. 
 \\ \notag
& \left. \phantom{ =\ } +\sum_{i=0}^{\ell-1} (-1)^{i+1} \Omega_{\alpha,0;\alpha,i} \Omega_{\beta,\ell-1-i;\beta,p} \right) \, ,
\end{align}
while in the case $\alpha=\beta$ can be rewritten as
\begin{align} \label{eq:def-Omega-a}
& \widehat{\r_\ell z^\ell}[u]. \Omega _{\alpha,0;\alpha,p}  = 
\frac{\hbar}{2} (\r_\ell)_{\alpha\alpha} \sum_{i=0}^{\ell-1} (-1)^{i+1}
\frac{\d^2 \Omega_{\alpha,0;\alpha,p}}{\d q_{\alpha,i} \d q_{\alpha,\ell-1-i}}
\\ \notag
& + (\r_\ell)_{\alpha\alpha}\left( \Omega_{\alpha,\ell;\alpha,p} + \Omega_{\alpha,0;\alpha,p+\ell} 
+\sum_{i=0}^{\ell-1} (-1)^{i+1} \Omega_{\alpha,0;\alpha,i} \Omega_{\alpha,\ell-1-i;\alpha,p} \right) 
\\ \notag
& + (-1)^{\ell-1}O \Omega_{\alpha,0;\alpha,p}\, ,
\end{align}
where $O$ is an operator defined as $O=O_1+O_2$, where
\begin{align}\label{eq:O-1-2}
& O_1:=  -\sum_{n=0}^\infty \left(
\sum_\beta (\r_\ell)_{\alpha\beta} \d_x^n \Omega_{\beta,\ell;\beta,0}
+ (-1)^{\ell-1} (\r_\ell)_{\alpha\un} (n+1) \d_x^n \Omega_{\alpha,\ell;\alpha,0} 
\right. \\ \notag & \left.
+ \sum_\beta (\r_\ell)_{\alpha\beta}
\sum_{i=0}^{\ell-1} (-1)^{i+1} \sum_{k=0}^n 
\binom{n+1}{k} \d_x^k \Omega_{\alpha,i;\alpha,0}
\d_x^{n-k} \Omega_{\beta,\ell-1-i;\beta,0} \right) \frac{\d}{\d u_{\alpha,n}} \,,
\\ \notag & 
O_2:=  -\sum_{n=0}^\infty \left(\frac{\hbar}{2} (\r_\ell)_{\alpha\alpha} 
\sum_{i=0}^{\ell-1} (-1)^{i+1} \d_x^{n+1} \Omega_{\alpha,0;\alpha,i;\alpha,\ell-1-i}
\right) \frac{\d}{\d u_{\alpha,n}} \,.
\end{align}
We have denoted $\Omega_{\alpha,p;\beta,q;\gamma,r} = \frac{\partial}{\partial q_{\gamma,r}} \Omega_{\alpha,p;\beta,q}$ and used the identities
\begin{equation}
u_\alpha = \Omega_{\alpha,0;\alpha,0}, \quad
\frac{\partial u_{\alpha,n}}{\partial q_{\alpha,i}} = \partial_x^{n+1} \Omega_{\alpha,i;\alpha,0}. \notag
\end{equation}

\begin{theorem} These deformations coincide with the ones we obtain taking the residue of the Sato-Wilson equation.
\end{theorem}

\begin{proof} Observe that in the case $\alpha\not=\beta$ the formulas given by Equations~\eqref{eq:def-Omega-a-b} and \eqref{eq:R-ResNewSatoab} do coincide, up to multiplication by $(-1)^{\ell-1}$, which is exactly the sign we omitted in Equation~\eqref{eq:first-r}. Therefore we only need to consider the case  $\alpha=\beta$, i.e. to show that the complicated deformation formulas  given, in the two approaches, by the Equations~\eqref{eq:R-aa} and~\eqref{eq:def-Omega-a}, are equivalent.

First of all note that the second line in~\eqref{eq:def-Omega-a} can be written
\begin{align}
& (\r_\ell)_{\alpha\alpha}\left( \Omega_{\alpha,\ell;\alpha,p} + \Omega_{\alpha,0;\alpha,p+\ell} 
+\sum_{i=0}^{\ell-1} (-1)^{i+1} \Omega_{\alpha,0;\alpha,i} \Omega_{\alpha,\ell-1-i;\alpha,p} \right) 
\\ \notag
& = \hbar \Res_\d \left(\frac{1}{\sqrt\hbar} (\r_\ell)_{\alpha\alpha} a_{p+\ell} L_\alpha^{p+\ell+\frac12}
+\frac1{\sqrt\hbar} (\r_\ell)_{\alpha\alpha}
\Omega_{\alpha,\ell;\alpha,p}L_\alpha^{-\frac12} 
\right. \\ \notag
& \left. +\frac1{\sqrt\hbar} (\r_\ell)_{\alpha\alpha}
\sum_{i=0}^{\ell-1}(-1)^{i+1} a_i \Omega_{\alpha,\ell-1-i;\alpha,p}L_\alpha^{i+\frac12}
\right)\, ,
\end{align}
reproducing lines 4, 5 and 7 in~\eqref{eq:R-aa}.

The operator $O_1$ has a nontrivial commutator with $\d_x$ given by
\begin{align}\label{eq:O1dx}
[O_1,\d_x]= & - \left( (-1)^{\ell-1} (\r_\ell)_{\alpha\un} \frac{\d}{\d q_{\alpha,\ell}} 
\right. \\ \notag & \left.
+ 
\sum_\beta (\r_\ell)_{\alpha\beta} \sum_{i=1}^{\ell-1} (-1)^{i+1} \Omega_{\beta,0;\beta,\ell-1-i} \frac{\d}{\d q_{\alpha,i}} \right),
\end{align}
where we have used the identities
\begin{equation}
[ \frac{\partial}{\partial u_{\alpha,p}} , \partial_x ] = \frac{\partial}{\partial u_{\alpha,n-1}} , \quad
\frac{\partial}{\partial q_{\alpha,i}} = \sum_{n\geq0} \partial_x^{n+1} \Omega_{\alpha,0;\alpha,i} \frac{\partial}{\partial u_{\alpha,n}}. \notag
\end{equation}

Recalling that
\begin{equation}
\Omega_{\alpha,0;\alpha,p} = \shb a_p \Res_\partial L_\alpha^{p+\frac12} \, ,
\end{equation}
this implies that 
\begin{align} \label{O1om}
& O_1(\Omega_{\alpha,0;\alpha,p}) = \sqrt\hbar a_p \Res_\d [O_1,L_\alpha^{p+\frac12}]
\\ \notag
& = \sqrt\hbar a_p \Res_\d \left( [O_1,L_\alpha^{\frac12}] L_\alpha^p + L_\alpha^{\frac12}[O_1,L_\alpha^{\frac12}] L_\alpha^{p-\frac12} +
\cdots + L_\alpha^p [O_1,L_\alpha^{\frac12}] \right) \, .
\end{align}
Here $[O_1,L_\alpha^{p+\frac12}]$ is a pseudo-differential operator in $\d_x$, whose coefficients are differential operators in $\frac{\d}{\d q_{\alpha,i}}$. When we take the residue of such an operator, we always mean the coefficient of $\d_x^{-1}$, which does not contain the derivatives $\frac{\d}{\d q_{\alpha,i}}$, as one can easily check. 

To proceed, we want to replace $[O_1,L_\alpha^{p+\frac12}]$ with a usual pseudo-differential operator, whose coefficients are just functions. That is indeed possible by using the operator $Y_1$ given by the formula
\begin{align}\label{eq:Y1}
Y_1:= & \left[O_1,L_\alpha^{\frac12}\right] + \left[L_\alpha^{\frac12},x\right]  (-1)^{\ell-1} (\r_\ell)_{\alpha\un} \frac{\d}{\d q_{\alpha,\ell}} 
\\ \notag
& + \sum_\beta (\r_\ell)_{\alpha\beta} \sum_{i=1}^{\ell-1} (-1)^{i+1} \left[ L_\alpha^{\frac12}, \hbar \frac{\d\log\tau}{\d q_{\beta,\ell-1-i}}\right] \frac{\d}{\d q_{\alpha,i}}
\\ \notag
&- \left[L_\alpha^{\frac12},x\right]  (-1)^{\ell-1} (\r_\ell)_{\alpha\un} \frac{a_\ell}{\sqrt\hbar}\left(L_\alpha^{\ell+\frac12}\right)_+
\\ \notag
& - \sum_\beta (\r_\ell)_{\alpha\beta} \sum_{i=1}^{\ell-1} (-1)^{i+1} \left[ L_\alpha^{\frac12}, \hbar \frac{\d\log\tau}{\d 
q_{\beta,\ell-1-i}}\right] \frac{a_i}{\sqrt\hbar} \left(L_\alpha^{i+\frac12}\right)_+
\, .
\end{align}
First, it follows from Equation~\eqref{eq:O1dx} that $Y_1$ is a pseudo-differential operator. Indeed, the sum of the commutators with $\d_x$ (considered as a factor in $L_\alpha^{\frac12}$) that emerge in the second and the third summands on the right hand side of Equation~\eqref{eq:Y1} is equal, with an opposite sign, to the commutator of $O_1$ and $\d_x$ given by Equation~\eqref{eq:O1dx}. Second, from the Lax equation~\eqref{Lax} it follows that~\eqref{O1om} can be rewritten as
\begin{align} \label{cY1}
& \sqrt\hbar a_p \Res_\d \left( [O_1,L_\alpha^{\frac12}] L_\alpha^p + L_\alpha^{\frac12}[O_1,L_\alpha^{\frac12}] L_\alpha^{p-\frac12} +
\cdots + L_\alpha^p [O_1,L_\alpha^{\frac12}] \right) 
\\ \notag
& = \sqrt\hbar a_p \Res_\d \left( \left( Y_1 L_\alpha^p + L_\alpha^{\frac12}Y_1 L_\alpha^{p-\frac12} +
\cdots + L_\alpha^p Y_1 \right) \right.
\\ \notag
&+ \left[L_\alpha^{p+\frac12},x\right]  (-1)^{\ell-1} (\r_\ell)_{\alpha\un} \frac{a_\ell}{\sqrt\hbar}\left(L_\alpha^{\ell+\frac12}\right)_+
\\ \notag
& \left. + \sum_\beta (\r_\ell)_{\alpha\beta} \sum_{i=1}^{\ell-1} (-1)^{i+1} \left[ L_\alpha^{p+\frac12}, \hbar \frac{\d\log\tau}{\d 
q_{\beta,\ell-1-i}}\right] \frac{a_i}{\sqrt\hbar} \left(L_\alpha^{i+\frac12}\right)_+ \right) \, . 
\end{align}

Observe that the last two lines give
\begin{align}
& \sqrt\hbar a_p \Res_\d \left( \left[L_\alpha^{p+\frac12},x\right]  (-1)^{\ell-1} (\r_\ell)_{\alpha\un} \frac{a_\ell}{\sqrt\hbar}\left(L_\alpha^{\ell+\frac12}\right)_+ \right.
\\ \notag
& \left. + \sum_\beta (\r_\ell)_{\alpha\beta} \sum_{i=1}^{\ell-1} (-1)^{i+1} \left[ L_\alpha^{p+\frac12}, \hbar \frac{\d\log\tau}{\d 
q_{\beta,\ell-1-i}}\right] \frac{a_i}{\sqrt\hbar} \left(L_\alpha^{i+\frac12}\right)_+ \right)  
\\ \notag
& =\sqrt\hbar a_p \Res_\d \left( \left[\left(L_\alpha^{p+\frac12}\right)_-,x\right]  (-1)^{\ell-1} (\r_\ell)_{\alpha\un} \frac{a_\ell}{\sqrt\hbar}\left(L_\alpha^{\ell+\frac12}\right) \right.
\\ \notag
& \left. + \sum_\beta (\r_\ell)_{\alpha\beta} \sum_{i=1}^{\ell-1} (-1)^{i+1} \left[ \left(L_\alpha^{p+\frac12}\right)_-, \hbar \frac{\d\log\tau}{\d 
q_{\beta,\ell-1-i}}\right] \frac{a_i}{\sqrt\hbar} \left(L_\alpha^{i+\frac12}\right) \right) \, ,
\end{align}
so in this way we identify other two summands in Equation~\eqref{eq:R-aa}.

The remaining terms of Equation~\eqref{eq:def-Omega-a} can be rewritten in the following way:
\begin{align} \label{cY2}
&\frac{\hbar}{2} (\r_\ell)_{\alpha\alpha} \sum_{i=0}^{\ell-1} (-1)^{i+1}
\frac{\d^2 \Omega_{\alpha,0;\alpha,p}}{\d q_{\alpha,i} \d q_{\alpha,\ell-1-i}}
+ O_2 \Omega_{\alpha,0;\alpha,p} 
\\ \notag
& = \sqrt\hbar a_p \Res_\d \left(  Y_2 L_\alpha^p + L_\alpha^{\frac12}Y_2 L_\alpha^{p-\frac12} +
\cdots + L_\alpha^p Y_2\right.
\\ \notag
& \left. + \hbar (\r_\ell)_{\alpha\alpha} 
\sum_{i=0}^{\ell-1} (-1)^{i+1}\sum_{a+b+c=2p-1} \left(L_\alpha^\frac12\right)^a \frac{\d L_\alpha^\frac12}{\d q_{\alpha,i}} 
\left(L_\alpha^\frac12\right)^b \frac{\d L_\alpha^\frac12}{\d q_{\alpha,i}} \left(L_\alpha^\frac12\right)^c\right) \, ,
\end{align}
where 
\begin{align}
& Y_2:=
\frac{\hbar}{2} (\r_\ell)_{\alpha\alpha} \sum_{i=0}^{\ell-1} (-1)^{i+1}
\frac{\d^2 \left(L_\alpha^\frac12 \right)}{\d q_{\alpha,i} \d q_{\alpha,\ell-1-i}}
\\ \notag &
\left[ -\sum_{n=0}^\infty \left(\frac{\hbar}{2} (\r_\ell)_{\alpha\alpha} 
\sum_{i=0}^{\ell-1} (-1)^{i+1} \d_x^{n+1} \Omega_{\alpha,0;\alpha,i;\alpha,\ell-1-i}
\right) \frac{\d}{\d u_{\alpha,n}}
, L_\alpha^\frac12 \right] 
\end{align}
Meanwhile, using the Lax equation~\eqref{Lax}, we see that the last term in~\eqref{cY2} is
\begin{align} \label{cY3}
& \sqrt\hbar a_p \Res_\d \hbar{(\r_\ell)_{\alpha\alpha}}
\sum_{i=0}^{\ell-1} (-1)^{i+1}
\sum_{
\begin{smallmatrix}
a+b+c\\
=2p-1
\end{smallmatrix}} \left(L_\alpha^\frac12\right)^a \frac{\d L_\alpha^\frac12}{\d q_{\alpha,i}} 
\left(L_\alpha^\frac12\right)^b \frac{\d L_\alpha^\frac12}{\d q_{\alpha,i}} \left(L_\alpha^\frac12\right)^c
\\ \notag &
= \sqrt\hbar a_p \Res_\d \left(  Y_3 L_\alpha^p + L_\alpha^{\frac12}Y_3 L_\alpha^{p-\frac12} +
\cdots + L_\alpha^p Y_3\right) 
\\ \notag & 
- \sqrt\hbar a_p \Res_\d \hbar{(\r_\ell)_{\alpha\alpha}} \sum_{i=0}^{\ell-1} (-1)^{i+1} \frac{\d \left(L_\alpha^{p+\frac12}\right)}{\d q_{\alpha,i}} 
\frac{a_{\ell-1-i}}{\sqrt\hbar}\left(L_\alpha^{\ell-i-\frac12}\right)_+\, .
\end{align}
The last line is equal to 
\begin{equation}
-\hbar \Res_\d 
{(\r_\ell)_{\alpha\alpha}} \sum_{i=0}^{\ell-1}(-1)^{i+1} a_i a_{p} 
\frac{\d \left(L_\alpha^{p+\frac{1}{2}}\right)_-}{\d q_{\alpha,\ell-i-1}} L_\alpha^{i+\frac{1}{2}}\, ,
\end{equation}
which corresponds to the last summand in Equation~\eqref{eq:R-aa} that we haven't yet reproduced, and $Y_3$ is given by the formula
\begin{equation}
Y_3:= \hbar{(\r_\ell)_{\alpha\alpha}} \sum_{i=0}^{\ell-1} (-1)^{i+1} \frac{\d \left(L_\alpha^{\frac12}\right)}{\d q_{\alpha,i}} 
\frac{a_{\ell-1-i}}{\sqrt\hbar}\left(L_\alpha^{\ell-i-\frac12}\right)_+\, .
\end{equation}

If we show that $Y:=Y_1+Y_2+Y_3$ solves~\eqref{eq:Y} then the contribution from the terms in~\eqref{cY1}, \eqref{cY2}, \eqref{cY3} that we have not matched yet, i.e. 
\begin{equation}
\shb a_p \Res_\partial \left( Y L_\alpha^{p} +L_\alpha^{\frac12}Y L_\alpha^{p-\frac12} +
\cdots + L_\alpha^p Y\right) 
\end{equation}
gives exactly the deformation term of~\eqref{YY125}.

So, in order to complete the proof of the theorem, it is sufficient to show that  
$Y:=Y_1+Y_2+Y_3$ is a solution of the Equation~\eqref{eq:Y}. This we do below, in Lemma~\ref{lem:solY}.
\end{proof}
 
\begin{lemma}\label{lem:solY}
The pseudo-differential operator $Y:=Y_1+Y_2+Y_3$ solves Equation~\eqref{eq:Y}.
\end{lemma}
\begin{proof}
First, observe that
\begin{align}
& L_\alpha^\frac12 Y_1 + Y_1 L_\alpha^\frac12 = 
\left[O_1,L_\alpha\right]
\\ \notag &
 + \left[L_\alpha,x\right]  (-1)^{\ell-1} (\r_\ell)_{\alpha\un} \frac{\d}{\d q_{\alpha,\ell}} 
+  \left[L_\alpha^\frac12,x\right]  (-1)^{\ell-1} (\r_\ell)_{\alpha\un} \frac{\d L_\alpha^\frac12}{\d q_{\alpha,\ell}} 
\\ \notag
& + \sum_\beta (\r_\ell)_{\alpha\beta} \sum_{i=1}^{\ell-1} (-1)^{i+1} \left[ L_\alpha, \hbar \frac{\d\log\tau}{\d q_{\beta,\ell-1-i}}\right] \frac{\d}{\d q_{\alpha,i}}
\\ \notag
& + \sum_\beta (\r_\ell)_{\alpha\beta} \sum_{i=1}^{\ell-1} (-1)^{i+1} \left[ L_\alpha^{\frac12}, \hbar \frac{\d\log\tau}{\d q_{\beta,\ell-1-i}}\right] \frac{\d L_\alpha^\frac12}{\d q_{\alpha,i}}
\\ \notag
&- \left[L_\alpha,x\right]  (-1)^{\ell-1} (\r_\ell)_{\alpha\un} \frac{a_\ell}{\sqrt\hbar}\left(L_\alpha^{\ell+\frac12}\right)_+
\\ \notag
&- \left[L_\alpha^{\frac12},x\right]  (-1)^{\ell-1} (\r_\ell)_{\alpha\un} \frac{a_\ell}{\sqrt\hbar}
\left[\left(L_\alpha^{\ell+\frac12}\right)_+, L_\alpha^{\frac12}\right]
\\ \notag
& - \sum_\beta (\r_\ell)_{\alpha\beta} \sum_{i=1}^{\ell-1} (-1)^{i+1} \left[ L_\alpha, \hbar \frac{\d\log\tau}{\d 
q_{\beta,\ell-1-i}}\right] \frac{a_i}{\sqrt\hbar} \left(L_\alpha^{i+\frac12}\right)_+
\\ \notag
& - \sum_\beta (\r_\ell)_{\alpha\beta} \sum_{i=1}^{\ell-1} (-1)^{i+1} \left[ L_\alpha^{\frac12}, \hbar \frac{\d\log\tau}{\d 
q_{\beta,\ell-1-i}}\right] \frac{a_i}{\sqrt\hbar}\left[ \left(L_\alpha^{i+\frac12}\right)_+, L_\alpha^{\frac12}\right]
\, .
\end{align}
Using the Lax equation~\eqref{Lax-L-a} and the explicit formulas for $O_1$ and $L_\alpha$ (Equations~\eqref{eq:O-1-2} and~\eqref{L-a}, respectively), we can rewrite this as
\begin{align}
& L_\alpha^\frac12 Y_1 + Y_1 L_\alpha^\frac12 = 
\left[L_\alpha,x\right]  (-1)^{\ell-1} (\r_\ell)_{\alpha\un} \frac{\d}{\d q_{\alpha,\ell}} 
\\ \notag
& + \sum_\beta (\r_\ell)_{\alpha\beta} \sum_{i=1}^{\ell-1} (-1)^{i+1} \left[ L_\alpha, \hbar \frac{\d\log\tau}{\d q_{\beta,\ell-1-i}}\right] \frac{\d}{\d q_{\alpha,i}}
\\ \notag
&- \left[L_\alpha,x\right]  (-1)^{\ell-1} (\r_\ell)_{\alpha\un} \frac{a_\ell}{\sqrt\hbar}\left(L_\alpha^{\ell+\frac12}\right)_+
\\ \notag
& - \sum_\beta (\r_\ell)_{\alpha\beta} \sum_{i=1}^{\ell-1} (-1)^{i+1} \left[ L_\alpha, \hbar \frac{\d\log\tau}{\d 
q_{\beta,\ell-1-i}}\right] \frac{a_i}{\sqrt\hbar} \left(L_\alpha^{i+\frac12}\right)_+
\\ \notag
&  - 2\cdot \left(
\sum_\beta (\r_\ell)_{\alpha\beta} \Omega_{\beta,\ell;\beta,0}
+ (-1)^{\ell-1} (\r_\ell)_{\alpha\un} \Omega_{\alpha,\ell;\alpha,0} 
\right. 
\\ \notag & \left.
+ \sum_\beta (\r_\ell)_{\alpha\beta}
\sum_{i=0}^{\ell-1} (-1)^{i+1} 
\Omega_{\alpha,i;\alpha,0}
\Omega_{\beta,\ell-1-i;\beta,0} \right) 
\\ \notag 
&  - 2 \hbar \left( (-1)^{\ell-1} (\r_\ell)_{\alpha\un} \frac{\d}{\d q_{\alpha,\ell}} 
\right. \\ \notag & \left.
+
\sum_\beta (\r_\ell)_{\alpha\beta} \sum_{i=1}^{\ell-1} (-1)^{i+1} \Omega_{\beta,0;\beta,\ell-1-i} \frac{\d}{\d q_{\alpha,i}} \right) \d_x 
\\ \notag &
- \hbar \sum_\beta (\r_\ell)_{\alpha\beta} 
\sum_{i=1}^{\ell-1} (-1)^{i+1} \d_x \Omega_{\beta,0;\beta,\ell-1-i} \frac{\d}{\d q_{\alpha,i}}
\\ \notag
&  = 
- \left[L_\alpha,x\right]  (-1)^{\ell-1} (\r_\ell)_{\alpha\un} \frac{a_\ell}{\sqrt\hbar}\left(L_\alpha^{\ell+\frac12}\right)_+
\\ \notag
& - \sum_\beta (\r_\ell)_{\alpha\beta} \sum_{i=1}^{\ell-1} (-1)^{i+1} \left[ L_\alpha, \hbar \frac{\d\log\tau}{\d 
q_{\beta,\ell-1-i}}\right] \frac{a_i}{\sqrt\hbar} \left(L_\alpha^{i+\frac12}\right)_+
\\ \notag
&  - 2\cdot \left(
\sum_\beta (\r_\ell)_{\alpha\beta} \Omega_{\beta,\ell;\beta,0}
+ (-1)^{\ell-1} (\r_\ell)_{\alpha\un} \Omega_{\alpha,\ell;\alpha,0} 
\right. 
\\ \notag & \left.
+ \sum_\beta (\r_\ell)_{\alpha\beta}
\sum_{i=0}^{\ell-1} (-1)^{i+1} 
\Omega_{\alpha,i;\alpha,0}
\Omega_{\beta,\ell-1-i;\beta,0} \right) \, .
\end{align}

In order to match this expression to the right hand side of the Equation~\eqref{eq:Y}, we note that
\begin{align}
\left(\frac{a_\ell}{\sqrt\hbar} [L_\alpha,x] L_\alpha^{\ell+\frac12}\right)_+
& =
2 \Omega_{\alpha,0;\alpha,\ell}+ \frac{a_\ell}{\sqrt\hbar} [L_\alpha,x] \left(L_\alpha^{\ell+\frac12}\right)_+ \, ,
\\ \notag
\left(\frac{a_i}{\sqrt\hbar} \left[ L_\alpha, \frac{\hbar \partial \log \tau}{\partial  q_{\beta,j}}\right] L_\alpha^{i+\frac12} \right)_+
& = 2\Omega_{\alpha,i;\alpha,0}
\Omega_{\beta,j;\beta,0}
\\ \notag & \phantom{ = } +
\left[ L_\alpha, \frac{\hbar\d\log\tau}{\d 
q_{\beta,j}}\right] \frac{a_i}{\sqrt\hbar} \left(L_\alpha^{i+\frac12}\right)_+ \, ,
\\ \notag 
\left({\sqrt\hbar}
\left[ L_\alpha, \frac{\partial\log\tau}{\partial q_{\beta,\ell}}\right] L_\alpha^{-\frac12}
\right)_+ 
& = 2 \Omega_{\beta,\ell;\beta,0} \, .
\end{align}

Then we observe that
\begin{align}
& L_\alpha^\frac12 (Y_2+Y_3) + (Y_2+Y_3) L_\alpha^\frac12 = 
\\ \notag & 
\frac{\hbar}{2} (\r_\ell)_{\alpha\alpha} \sum_{i=0}^{\ell-1} (-1)^{i+1}
\left( \frac{\d^2 \left(L_\alpha^\frac12 \right)}{\d q_{\alpha,i} \d q_{\alpha,\ell-1-i}} L_\alpha^\frac12 
+ L_\alpha^\frac12 \frac{\d^2 \left(L_\alpha^\frac12 \right)}{\d q_{\alpha,i} \d q_{\alpha,\ell-1-i}} \right)
\\ \notag &
- 2 \left(\frac{\hbar}{2} (\r_\ell)_{\alpha\alpha} 
\sum_{i=0}^{\ell-1} (-1)^{i+1} \d_x \Omega_{\alpha,0;\alpha,i;\alpha,\ell-1-i}
\right) 
\\ \notag &
+\hbar{(\r_\ell)_{\alpha\alpha}} \sum_{i=0}^{\ell-1} (-1)^{i+1} \frac{\d \left(L_\alpha^\frac12\right)}{\d q_{\alpha,i}} 
\frac{a_{\ell-1-i}}{\sqrt\hbar}\left[ \left(L_\alpha^{\ell-i-\frac12}\right)_+, L_\alpha^\frac12\right]
\\ \notag &
+\hbar{(\r_\ell)_{\alpha\alpha}} \sum_{i=0}^{\ell-1} (-1)^{i+1} \frac{\d L_\alpha}{\d q_{\alpha,i}} 
\frac{a_{\ell-1-i}}{\sqrt\hbar}\left(L_\alpha^{\ell-i-\frac12}\right)_+\, .
\end{align}
The sum of the first three summands on the right hand side of this equation is equal to zero. The last summand is a differential operator, that is, has no negative part, and the Lax equations imply that  
\begin{align}
& \hbar{(\r_\ell)_{\alpha\alpha}} \sum_{i=0}^{\ell-1} (-1)^{i+1} \frac{\d L_\alpha}{\d q_{\alpha,i}} 
\frac{a_{\ell-1-i}}{\sqrt\hbar}\left(L_\alpha^{\ell-i-\frac12}\right)_+
\\ \notag &
= \left(
{(\r_\ell)_{\alpha\alpha}} 
\sum_{i=0}^{\ell-1}(-1)^{i+1} a_i a_{\ell-1-i} \left[ L_\alpha, \left(L_\alpha^{\ell-i-\frac{1}{2}}\right)_-\right] L_\alpha^{i+\frac{1}{2}}
\right)_+ \, ,
\end{align}
which is the only remaining summand on the right hand side of Equation~\eqref{eq:Y}. 

Thus we see that $Y_1+Y_2+Y_3$ solves  Equation~\eqref{eq:Y}.
\end{proof}

\subsection{The $S$-deformations} We recall the formula for the $S$-de\-for\-ma\-tion of $\Omega_{\alpha,p;\beta,q}$ obtained in~\cite[Thm. 9]{BurPosSha1}:

\begin{align} \label{eq:def-Omega-low}
& \sum_{\ell=1}^\infty \widehat{\s_\ell z^{-\ell}}[u]. \Omega _{\alpha,p;\beta,q} = 
\sum_{1\leq\ell\leq p}(\s_\ell)^{\mu}_{\alpha}\Omega_{\mu,p-\ell;\beta,q}+\sum_{1\leq\ell\leq q}
\Omega_{\alpha,p;\mu,q-\ell} (\s_\ell)^{\mu}_{\beta} 
\\ \notag
& + (-1)^p(\s_{p+q+1})_{\alpha\beta}-\sum_{\gamma}\frac{\partial\Omega_{\alpha,p;\beta,q}}{\partial u_{\gamma,0}} (\s_1)_{\gamma,\un}.
\end{align}
As before indices are raised and lowered according to the rule~\eqref{raising}.

In the special case of several copies of KdV, and for $p=0$, this formula can be simplified. Indeed, if $\alpha\not=\beta$, then
\begin{equation}\label{eq:def-BPS-S-ab}
\sum_{\ell=1}^\infty \widehat{\s_\ell z^{-\ell}}[u]. \Omega _{\alpha,0;\beta,p} = \sum_{\ell=1}^p (\s_\ell)_{\alpha\beta} \Omega_{\alpha,0;\alpha,p-\ell} + (\s_{p+1})_{\alpha\beta}.
\end{equation}
In the case $\alpha=\beta$, we have
\begin{align} \label{eq:DefOmegaSaa}
\sum_{\ell=1}^\infty \widehat{\s_\ell z^{-\ell}}[u]. \Omega _{\alpha,0;\alpha,p} = & \sum_{\ell=1}^p (\s_\ell)_{\alpha\alpha} \Omega_{\alpha,0;\alpha,p-\ell} + (\s_{p+1})_{\alpha\alpha} 
\\ \notag & 
- (\s_1)_{\alpha,\un}\Omega_{\alpha,0;\alpha,p-1} .
\end{align}

\begin{theorem} These deformations coincide with the ones we obtain taking the residue of the Sato-Wilson equation.
\end{theorem}

\begin{proof} Indeed, the right hand side of Equation~\eqref{eq:def-BPS-S-ab} is equal to the right hand side of Equation~\eqref{eq:ResNewSatoab}. From Lemma~\ref{lem:ResLtilde} and Equation~\eqref{eq:LasSDef} it follows that the $\epsilon$-term on the right hand side of Equation~\eqref{eq:ResDefS} multiplied by $\hbar$ is equal to the right hand side of Equation~\eqref{eq:DefOmegaSaa}.
\end{proof}


\end{document}